\documentclass[a4paper,12pt]{article}
\usepackage{amsmath}
\usepackage{amssymb}
\usepackage{tabularx}
\usepackage{enumerate}
\usepackage{graphicx}
\usepackage{subfigure}
\usepackage{color}
\usepackage[pagewise]{lineno}
\usepackage{longtable}
\usepackage{epstopdf}
\usepackage{float}
\usepackage{setspace}

\usepackage[justification=centering]{caption}

\usepackage{hyperref}
\hypersetup{
colorlinks=true,
linkcolor=red,
anchorcolor=red,
citecolor=red}

\usepackage{graphicx}
\usepackage{subfigure}
\newtheorem{thm}{Theorem}[section]
\newtheorem{lem}[thm]{Lemma}

\newtheorem{exm}{Example}[section]

\newtheorem{pppp}{Proof}

\newcommand{\qed}{\hspace{1em}\mbox{\raisebox{0.65ex}{\fbox{}}}}

\numberwithin{equation}{section}

\newcommand{\be}{\begin{equation}}
\newcommand{\ee}{\end{equation}}
\newcommand\bes{\begin{eqnarray}}
\newcommand\ees{\end{eqnarray}}
\newcommand{\bess}{\begin{eqnarray*}}
\newcommand{\eess}{\end{eqnarray*}}

\newcommand{\bpf}{{\bf Proof:\ \ }}
\newcommand{\epf}{\mbox{}\hfill $\Box$}

\begin{document}
\thispagestyle{empty}

\title{Interaction between harvesting intervention and birth perturbation in an age-structured model\thanks{ The first author is supported by the Natural Science Foundation of Jiangsu Province, PR China (No. BK20220553), the second author is supported by the National Natural Science Foundation of China (No. 12271470) and Carlos Alberto Santos acknowledges the support of CNPq/Brazil Proc. $N^o$ $311562/2020-5$, and FAPDF grant 00193.00001133/2021-80.}}

\date{\empty}
\author{Haiyan Xu$^1$, Zhigui Lin$^1$\thanks{Corresponding author. Email: zglin@yzu.edu.cn (Z. Lin).} and Carlos Alberto Santos$^2$ \\
{\small $^1$ School of Mathematical Science, Yangzhou University, Yangzhou 225002, China}\\
{\small $^2$ Department of Mathematics, University of Brasilia, BR-70910900 Brasilia, DF, Brazil}
}

 \maketitle

\begin{quote}
\noindent {\bf Abstract.}
 An age-structured fish model with birth and harvesting pulses is established, where birth pulses are responsible for increasing the amount of fish due to the constant multiple placement of juveniles, and harvesting pulses  describe the decrease of fish due to fishing activities. The principal eigenvalue as a threshold value depending on the harvesting and birth intensity is firstly investigated by three different ways. The asymptotic behavior of the population is fully investigated and sufficient conditions for the species to be extinct or persist are given. Numerical simulations suggest that interaction between negative harvesting intervention and positive birth perturbation  decides extinction and persistence of the species. It is also shown that perturbation timing plays an important role.
 \medskip


\end{quote}

\section{Introduction}

This paper introduces a juvenile-adult fish model with birth pulse (constant multiple placement of juvenile fish) exerting on juveniles and harvesting pulses (proportional capturing of fish) imposed on whole species, which reads as
\begin{eqnarray}
\left\{
\begin{array}{lll}
u_{1t}=d_1u_{1xx}+{b(t)}u_2-(a(t)+m_1(t))u_1-\alpha_1(t)u_1^2,  &(t,x)\in \Omega_{\rho,\tau},\\[1mm]
u_{2t}=d_2u_{2xx}+{a(t)}u_1-m_2(t)u_2-\alpha_2(t)u_2^2, & (t,x)\in \Omega_{\rho,\tau},\\[1mm]
u_1((n\tau)^+,x)=(1+\delta)u_1(n\tau,x), & x\in \overline \Omega, \\[1mm]
u_1(((n+\rho)\tau)^+,x)=(1-\theta)u_1((n+\rho)\tau,x), & x\in \overline \Omega, \\[1mm]
u_2(((n+\rho)\tau)^+,x)=(1-\theta)u_2((n+\rho)\tau,x), & x\in \overline \Omega, \\[1mm]
u_1(t,x)=u_2(t,x)=0, & t>0,\,\,x \in \partial \Omega, \\[1mm]
u_1(0,x)=u_{1,0}(x), u_2(0,x)=u_{2,0}(x),&x\in \overline \Omega,\ \\[1mm]
\end{array} \right.
\label{a01}
\end{eqnarray}
where $\Omega\in R^N$($N \geq 1 $) is a connected and bounded domain with smooth boundary $\partial \Omega$, and
$$\Omega_{\rho,\tau}:=\Omega_{\rho,\tau}^n :=\{(t,x):t\in((n\tau)^+,(n+\rho)\tau]\cup (((n+\rho)\tau)^+,(n+1)\tau],\,x \in \Omega\}$$
 for each $n=0,1,2,\cdots,$ and for any $0<\rho<1$ and $\tau>0$ given. The functions $u_1(t,x)$ and $u_2(t,x)$ represent the density of juvenile and adult fish at time $t$ and space $x$, $d_1$ and $d_2$ are the diffusion rate of juveniles and adults, respectively. Spatial-independent functions $m_1(t)$, $m_2(t)$, $b(t)$, $a(t)$, $\alpha_1(t)$ and $\alpha_2(t) \in C^{p/2, p}([0,\tau]\times \overline \Omega)$($0<p \leq 1$) are positive and $\tau-$periodic. We will denote the maximum and minimum of them by superscript such as $m_1(t)$ by $m_1^M$ and $m_1^m$, respectively. More specifically, the functions $m_1(t)$ and $m_2(t)$ depict the death rate of juveniles and adults, while $b(t)$ means the reproduction rate of adults, and $a(t)$ denotes the rate that juvenile fish mature into adult fish. $\alpha_1(t)$ and $\alpha_2(t)$ are coefficients of competition between juvenile and adult, respectively. Initial functions $u_{i,0}(x)$ satisfies $u_{i,0}(x)\in C^2(\overline \Omega)$, $u_{i,0}(x)\geq, \not\equiv 0$ for $x\in \overline \Omega$ and $u_{i,0}(x)=0$ for $x\in \partial \Omega$ with i=1,2.

The multiple impulses are introduced in the system, and the habitat $\Omega_{\rho, \tau}$, defined as above, shows that,
in the successive  stage $(n\tau,(n+\rho)\tau]$, the fish will grow and disperse, and the initial value in this stage is $u_i((n\tau)^+,x)$ because of impulse, while in the successive stage $((n+\rho)\tau, (n+1)\tau]$, the fish  grows and spreads with the new initial value  $u_i(((n+\rho)\tau)^+,x)$. The function $(1+\delta)u_1$ indicates constant multiple placement for juvenile fish with intensity $\delta>0$ , and  such pulse occurs at every time $t=n\tau$. The harvesting control is imposed to juveniles and adults at every time $t=(n+\rho)\tau$, the impulsive function is $(1-\theta)u_i$ with $i=1, 2$ and $\theta (0<\theta<1)$ represents harvesting control intensity.

On condition that $\delta=\theta=0$, problem \eqref{a01} is a natural extension of classical Fisher's equation that has been considered in \cite{HZ,CCM}, where a novel numerical method has been developed to calculate the spreading speed in such a juvenile-adult system with temporal variable environment. A nonlocal-delayed and age-structured model was set up by Fang et al. \cite{FL}, who indicates that invasion speed coincides with minimal speed of travelling wave fronts and their corresponding conclusion was discussed numerically. Subsequently, Fang, Gourley and Lou \cite{FG} put forward an age-structured system of competition solely within age classes, and presented results about small delays and linear stability analysis on the boundary equilibria that were proved by overcoming strong coupling. A stage-structured patch model  with stage duration  in heterogeneous environment was studied in \cite{AL}. The importance of age-structure was  strengthened and  minimal patch size and travelling wave speed were also presented. Besides age-structure, local dispersal, reproduction and mortality \cite{OL2001}, recently the distribution of species related to external perturbations has attracted more attention in reaction-diffusion model, see \cite{LYG2019, MLP2021, LL} for one-single impulsive model, \cite{ZW,LZ2009} for two-species competition model, \cite{XLZ} for cooperative model, \cite{LZY2011} for prey-predator model and \cite{Y} for patch model. There are also rich results corresponding to pulse intervention in epidemic model, such as SIR (Susceptible-infected-recovery) model \cite{SJ}, SIRS (Susceptible-infected-recovery-susceptible) model \cite{BZ2020}, SEIR (Susceptible-epidemic-infected-recovery) model \cite{D,DA}, Zika model \cite{LZ} and WNv (West Nile virus) model \cite{PL}.

Problem \eqref{a01} with $\delta=0$ and $0<\theta<1$ has been recently discussed in \cite{XLZ} with harvesting pulse acting just on adults. Here we  introduce multiple pulse perturbations with $\delta>0$ and $0<\theta<1$ in the model \eqref{a01} to consider an age-structured fish system. Since fish is usually harvested in autumn, so the placement of juvenile fish in spring is necessary for sustainable development of fishery resources.

There are some difficulties arising from multiple perturbations in such an age-structured system. Whether does the principal eigenvalue characterized by both pulses exist and how can we give a more visualized method to show the existence of the principal eigenvalue for an eigenvalue problem associate to  \eqref{a01}?  Moreover, how do harvesting pulse $\theta$, birth pulse $\delta$ and timing pulse  $\tau$ alter the dynamics of the species? All of these are worth researching.

This paper is organized as follows. The global existence and uniqueness of the solution are shown in Section 2. Section 3 presents three ways to investigate the principal eigenvalue in different cases: the theoretical result based on Krein-Rutman Theorem  for problem \eqref{a01} with multiple pulses in heterogeneous environment, the separation variables combining with fundamental solution matrix  for constant coefficient cooperative model with multiple pulses, and the direct calculation for one-single system with two pulses. Section 4 is devoted to dynamics of the solution, and numerical approximations involving harvesting intensity, birth intensity and timing of pulses are exhibited in Section 5.

\section{Existence, uniqueness and estimates}

Global existence and uniqueness of solutions to problem \eqref{a01} can be obtained by the bootstrap method. For $n=0$, birth pulse takes place at time $t=0$. Then the solution $(u_1(t,x),u_2(t,x))$ satisfies problem \eqref{a01} with a new initial value $(u_1(0^+, x),u_2(0,x))$ over the time interval $(0^+,\rho\tau]$. Recalling that $u_{i,0}(x)\in [C^2(\overline \Omega)]^2$, we can deduce that the new initial value of $u_1(0^+,x)=(1+\delta)u_0(x)\in C^2(\overline \Omega)$.
Hence, it follows from the classical theory of partial differential equation \cite{Lie96} that the unique solution $(u_1(t,x),u_2(t,x)) \in {C^{1,2}((0,\rho\tau]\times \overline \Omega)}^2$ to the problem \eqref{a01} for $t\in(0^+,\rho\tau]$. Similarly, $u_i((\rho\tau)^+,x)=(1-\theta)u_i(\rho\tau,x)$ becomes a new initial value for $t\in((\rho\tau)^+,\tau]$, which also belongs to $C^2(\overline \Omega)$ so that there exists a unique solution $(u_1(t,x),u_2(t,x))\in [C^{1,2}((\rho\tau,\tau]\times \overline \Omega)]^2$ of the problem \eqref{a01} for $t\in((\rho\tau)^+,\tau]$ and so on.

Besides, $(u_1(t,x),u_2(t,x))\leq(w(t),w(t))$ with $w(t)$ satisfying
\begin{eqnarray*}
\left\{
\begin{array}{lll}
w_{t}=\alpha w-\beta w^2, & t\in(0^+,\tau],  \\
w(0^+)=(1+\delta) w(0),\\
w_i(0)=||u_{i,0}(x)||,
\end{array} \right.
\end{eqnarray*}
where $\alpha=\max\{b^M,a^M\}$ and $\beta=\max\{\alpha_1^m,\alpha_2^m\}$. It is standard to verify that $$w(t)=\frac{\alpha(1+\delta)||u_{i,0}(x)||(\beta e^{\alpha t}-\beta)+\alpha e^{\alpha t}}{\alpha(1+\delta)||u_{i,0}(x)||}\leq(1+\delta)M_1,$$
with $M_1$ independent of $\delta$, so that $u_{1,2}(t,x)\leq\max \{||u_{i,0}(x)||,(1+\delta)M_1\}$ for $(t,x)\in[0,\tau]\times\overline\Omega$.

 For $n=1$ and by the same procedures, we find that problem \eqref{a01} admits a unique solution $(u_1,u_2)$ for $t\in(\tau^+,2\tau]$, $(u_1,u_2)(t,x)\in {C^{1,2}((\tau,\rho\tau+\tau]\times \overline \Omega)}^2\cup[C^{1,2}((\rho\tau+\tau,2\tau]\times \overline \Omega)]^2$ and $(u_1(t,x),u_2(t,x))\leq(w(t),w(t))$ with $w(t)$ satisfying
\begin{eqnarray*}
\left\{
\begin{array}{lll}
w_{t}=\alpha w-\beta w^2, & t\in(\tau^+,2\tau],  \\
w(\tau^+)=(1+\delta) w(\tau),\\
w(\tau)=(1+\delta)M_1,
\end{array} \right.
\end{eqnarray*}
that is,
 $$w(t)=\frac{\alpha(1+\delta)^2M_1e^{-\alpha \tau}+\alpha e^{\alpha t}}{\beta(e^{\alpha t}-e^{\alpha \tau})(1+\delta)^2M_1e^{-\alpha\tau}}\leq(1+\delta)^2M_2$$
with $M_2(\geq M_1)$ independent of $\delta$ so that $u_{1}(t,x),u_{2}(t,x)\leq (1+\delta)^2M_2$ for $(t,x)\in(\tau,2\tau]\times\overline\Omega$.

Step by step, we can find by Zorn's Lemma a positive constant $M$ independent of $\delta$ and a maximum time $T_{\max}$  such that the solution $(u_1,u_2)$ exists and is unique in $t\in[0,T_{\max})$, and satisfies
$$u_{i}(t,x)\leq \{||u_{i,0}(x)||,(1+\delta)^{[\frac{T_{\max}}{\tau}]+1}M\},\,(t,x)\in[0,T_{\max})\times\overline\Omega,\,i=1,2.$$
By standard continuous extension method, we have that $T_{\max}=\infty$. Therefore, problem \eqref{a01} admits a unique solution $(u_1,u_2)$ for $t\in(0,\infty)$. Besides,
\begin{align*}
(u_1(t,x),u_2(t,x))\in & [PC^{1,2}((0,+\infty)\times \overline \Omega)]^2\\
:=&\Big\{\cap^{\infty}_{n=0}[C^{1,2}((n\tau, (n+\rho)\tau]\times\overline \Omega)\cap C^{1,2}(((n+\rho)\tau, (n+1)\tau]\times\overline \Omega)]\Big\}^2
\end{align*}
and $u_{i}(t,x)\leq \{||u_{i,0}(x)||,(1+\delta)^{[\frac{t}{\tau}]+1}M\}$ for $i=1,2$ and $(t,x)\in [0,\infty)\times\overline \Omega.$

\section{\bf The principal eigenvalue}
As a key indicator to show the long-time behavior of solution, an eigenvalue problem is firstly introduced, and the existence of an principal eigenvalue and its properties depending on the birth pulse and harvesting pulse are shown below.

We first linearize the initial-boundary problem \eqref{a01} about $(0,0)$ and then consider the following periodic eigenvalue problem
\begin{eqnarray}
\left\{
\begin{aligned}
&\phi_{t}=d_1\phi_{xx}+{b(t)}\psi-(a(t)+m_1(t))\phi+\mu\phi, &&t\in(0^+,\rho\tau]\cup ((\rho\tau)^+,\tau],\,\,x \in \Omega,  \\
&\psi_{t}=d_2\psi_{xx}+{a(t)}\phi-m_2(t)\psi+\mu\psi, &&t\in(0^+,\rho\tau]\cup ((\rho\tau)^+,\tau],\,\,x \in \Omega,  \\
&\phi(0,x)=\phi(\tau, x), \,\psi(0,x)=\psi(\tau, x), && x\in \overline \Omega,\\
&\phi(0^+, x)=(1+\delta)\phi(0,x),&& x\in \overline \Omega,\\
&\phi((\rho\tau)^+, x)=(1-\theta)\phi(\rho\tau,x),&& x\in \overline \Omega,\\
&\psi((\rho\tau)^+, x)=(1-\theta)\psi(\rho\tau,x),&& x\in \overline \Omega,\\
&\phi(t,x)=\psi(t,x)=0, && t\in [0,\tau],\,\,x \in \partial \Omega.
\end{aligned} \right.
\label{a02}
\end{eqnarray}
 The existence of an principal eigenvalue $\mu$ for the problem \eqref{a02} without impulses can be guaranteed by Krein-Rutman theorem \cite{KR,LZZ,ZZ} directly.  Recently, the eigenvalue problem on a Banach space involving impulse can also be solved by using Poincar\'{e} map associated with periodic semiflow \cite{WZ}. Here, we will derive the existence of an principal eigenvalue from the problem itself, and we will overcome the difficulties introduced by multiple pulses through modifying the eigenfunction space and considering the following equivalent eigenvalue problem
\begin{eqnarray}
\left\{
\begin{array}{lll}
\xi_{1t}=d_1\xi_{1xx}+{b(t)}\eta_1-(a(t)+m_1(t))\xi_1+\mu\xi_1,& t\in(0,\rho\tau],\,\,x \in \Omega,  \\[1mm]
\xi_{2t}=d_1\xi_{2xx}+{b(t)}\eta_2-(a(t)+m_1(t))\xi_2+\mu\xi_2,& t\in(\rho\tau,\tau],\,\,x \in \Omega,  \\[1mm]
\eta_{1t}=d_2\eta_{1xx}+{a(t)}\xi_1-m_2(t)\eta_1+\mu\eta_1,& t\in(0,\rho\tau],\,\,x \in \Omega,  \\[1mm]
\eta_{2t}=d_2\eta_{2xx}+{a(t)}\xi_2-m_2(t)\eta_2+\mu\eta_2,& t\in(\rho\tau,\tau],\,\,x \in \Omega,  \\[1mm]
\xi_i(t,x)=\eta_i(t,x)=0,\,i=1,2, & t\in [0,\tau],\,\,x\in \partial\Omega,\\[1mm]
\xi_1(0,x)=(1+\delta) \xi_2(\tau,x),\, \xi_1(t,x)=\xi_1(\rho\tau,x), &t\in [\rho\tau,\tau],\, x\in \overline \Omega,\\[1mm]
\xi_2(\rho\tau,x)=(1-\theta) \xi_1(\rho\tau,x),\, \xi_2(t,x)=\xi_2(\rho\tau,x), &t\in [0,\rho\tau],\, x\in \overline \Omega,\\[1mm]
\eta_1(0,x)= \eta_2(\tau,x),\, \eta_1(t,x)=\eta_1(\rho\tau,x), &t\in [\rho\tau,\tau],\, x\in \overline \Omega,\\[1mm]
\eta_2(\rho\tau,x)=(1-\theta) \eta_1(\rho\tau,x),\, \eta_2(t,x)=\eta_2(\rho\tau,x), &t\in [0,\rho\tau],\, x\in \overline \Omega.
\end{array} \right.
\label{3.02}
\end{eqnarray}
In fact, we can take $(\phi,\psi)(t,x)=(\xi_1,\eta_1)(t,x)$ for $t\in (0,\rho\tau]$ and $(\phi,\psi)(t,x)=(\xi_2,\eta_2)(t,x)$ for $t\in (\rho\tau,\tau]$, respectively, $\phi(0^+,x)=\xi_1(0,x)$, $\phi(0,x)=\xi_2(\tau,x)$, $\phi((\rho\tau)^+,x)=\xi_2(\rho\tau,x)$, $\phi(\rho\tau,x)=\xi_1(\rho\tau,x)$, $\psi((\rho\tau)^+,x)=\eta_2(\rho\tau,x)$ and $\psi(\rho\tau,x)=\eta_1(\rho\tau,x)$ for $x\in \overline \Omega$.

Now let $W$ be the Banach space
\begin{eqnarray*}
\begin{array}{rll}
W=&{D}^{0,1}_0([0,\tau]\times \overline \Omega):=\big\{(\xi_1,\xi_2,\eta_1,\eta_2)\in [{C}^{0,1}([0, 1]\times \overline \Omega)]^4:\xi_1=\xi_2=\eta_1=\eta_2=0,\\  &(t,x)\in [0, \tau]\times \partial \Omega,
\ (\xi_1,\eta_1)(t,x)=(\xi_1,\eta_1)(\rho\tau,x), t\in [\rho\tau,\tau],\\&(\xi_2,\eta_2)(t,x)=(\xi_2,\eta_2)(\rho\tau,x),\, t\in [0,\rho\tau],\, x\in \overline \Omega, \\
&\xi_1(0,x)=(1+\delta) \xi_2(\tau,x),\ \xi_2(\rho\tau,x)=(1-\theta) \xi_1(\rho\tau,x),\\ &\eta_1(0,x)=\eta_2(\tau,x),\,\eta_2(\rho\tau,x)=(1-\theta)\eta_1(\rho\tau,x),\,  x\in \overline \Omega \big\}
\end{array}
\end{eqnarray*}
with the positive cone
${W^+}:=\textrm{closure}\{(\xi_1,\xi_2,\eta_1,\eta_2)\in W:\, \xi_i(t,x),\eta_i(t,x)\gg 0\ \forall (t,x)\in [0, \tau] \times \partial \Omega,\,i=1,2 \},$
and its nonempty interior
$\textrm{Int} (W^+)=\{(\xi_1,\xi_2,\eta_1,\eta_2)\in W:\ \xi_i(t,x),\eta_i(t,x)\\ \gg 0\ \forall (t,x)\in [0, \tau] \times \partial \Omega,\,i=1,2 \}.$

Let $M^*=1+\max_{[0,\tau]\times \overline \Omega} |a(t)+b(t)|+\ln (1/(1-\beta))$. For any given $(\xi_1,\xi_2,\eta_1,\eta_2)\in W$, the linear problem
\begin{eqnarray}
\left\{
\begin{array}{lll}
w_{1t}=d_1w_{1xx}+{b(t)}z_1-(a(t)+m_1(t)+M^*)w_1+\xi_1,& t\in(0,\rho\tau],\,\,x \in \Omega,  \\[1mm]
w_{2t}=d_1w_{2xx}+{b(t)}z_2-(a(t)+m_1(t)+M^*)w_2+\xi_2,& t\in(\rho\tau,\tau],\,\,x \in \Omega,  \\[1mm]
z_{1t}=d_2z_{1xx}+{a(t)}w_1-m_2(t)z_1-M^*+\eta_1,& t\in(0,\rho\tau],\,\,x \in \Omega,  \\[1mm]
z_{2t}=d_2z_{2xx}+{a(t)}w_2-m_2(t)z_2-M^*+\eta_2,& t\in(\rho\tau,\tau],\,\,x \in \Omega,  \\[1mm]
w_1(t,x)=z_1(t,x)=0, & t\in [0, \rho\tau],\,\,x\in \partial\Omega,\\[1mm]
w_2(t,x)=z_2(t,x)=0, & t\in [\rho\tau,\tau],\,\,x\in \partial\Omega,\\[1mm]
w_1(0,x)=(1+\delta) w_2(\tau,x),\, w_1(t,x)=w_1(\rho\tau,x), &t\in [\rho\tau,\tau],\, x\in \overline \Omega,\\[1mm]
w_2(\rho\tau,x)=(1-\theta) w_1(\rho\tau,x),\, w_2(t,x)=w_2(\rho\tau,x), &t\in [0,\rho\tau],\, x\in \overline \Omega,\\[1mm]
z_1(0,x)= z_2(\tau,x),\, z_1(t,x)=z_1(\rho\tau,x), &t\in [\rho\tau,\tau],\, x\in \overline \Omega,\\[1mm]
z_2(\rho\tau,x)=(1-\theta) z_1(\rho\tau,x),\, z_2(t,x)=z_2(\rho\tau,x), &t\in [0,\rho\tau],\, x\in \overline \Omega.
\end{array} \right.
\label{3.22}
\end{eqnarray}
admits a unique solution $(w_1,w_2,z_1,z_2)$ (\cite{Lie96}) satisfying $w_1,w_2,z_1, z_2 \in {C}^{(1+\alpha)/2, 1+\alpha}([0, \tau]\times \overline \Omega)\bigcap W$ for some $0<\alpha<1$, whence follows that the operator
$$\mathcal{A} (\xi_1,\xi_2,\eta_1,\eta_2) =(w_1,w_2,z_1,z_2)$$ is well defined. Since that the embedding ${C}^{(1+\alpha)/2,1+\alpha}\hookrightarrow {C}^{0,1}$ is compact, $\mathcal{A}$ is an  compact linear operator. Moreover, $\mathcal{A}$ is strongly positive in $W$. Therefore, it follows from Krein-Rutman theorem that there exist a unique $\sigma_1 := r(\mathcal{A}) > 0$ and a function  $(w_1,w_2,z_1,z_2)\in \textrm{Int} (W^+)$ such that $$\mathcal{A}(w_1,w_2,z_1,z_2) = \sigma_1(w_1,w_2,z_1,z_2),$$ then $\mu := 1/\sigma_1-M^*$
is the principal eigenvalue of \eqref{3.02} and the corresponding
eigenfunction pair $(\xi_i,\xi_2,\eta_1,\eta_2)$ is strongly positive. Coming back to problem \eqref{a02} with multiple pulses, we have the existence of the principal eigenvalue $\mu_1$ with positive eigenvalue function pair $(\phi,\psi)$ by equivalence.

\vspace{1mm}
Suppose all the coefficients in \eqref{a02} are constant. Then the first two equations in eigenvalue problem \eqref{a02} can be rewritten as
\begin{eqnarray}
\left\{
\begin{array}{lll}
f_{1t}=bf_2-(a+m_1+d_1\lambda_0)f_1+\mu f_1(t),& t\in(0^+,\rho\tau],\\[1mm]
f_{2t}=af_1-(m_2+d_2\lambda_0)f_2+\mu f_2,&t\in(0^+,\rho\tau]
\end{array} \right.
\label{c03}
\end{eqnarray}
by the method of separation variables, where $\lambda_0$ is the principal eigenvalue of $-\Delta$ satisfying $-\Psi_{xx}=\lambda_0\Psi$ for $x\in\Omega$ and $\Psi(x)=0$ for $x\in\partial\Omega$.

It follows from \eqref{c03} that
\begin{equation*}
\left(
\begin{array}{c}
f_{1}'(t)\\
f_{2}'(t)\\
\end{array}
\right)
=
\left(
\begin{array}{cc}
-a-m_1-d_1\lambda_0+\mu & b\\
a & -m_2-d_2\lambda_0+\mu\\
\end{array}
\right)
\left(
\begin{array}{c}
f_1(t)\\
f_2(t)\\
\end{array}
\right)
\triangleq M
\left(
\begin{array}{c}
f_1(t)\\
f_2(t)\\
\end{array}
\right),
\end{equation*}
where $f_i'(t) = f_{it}(t)$ for $t=1,2$.
The corresponding characteristic equation is $|M-\lambda E|=0$, and a direct calculation yields
\begin{equation}
\lambda_{1,2}=\mu+\frac{-[a+m_1+m_2+(d_1+d_2)\lambda_0]\underline{+}\sqrt{(a+m_1-m_2+d_1\lambda_0-d_2\lambda_0)^2+4ab}}{2}\triangleq \mu+c_{1,2}.
\label{bbb}
\end{equation}
 Without loss of generality we assume $c_1>c_2$, so $a+m_1+d_1\lambda_0+c_1=-(m_2+d_2\lambda_0+c_2)>0.$

Let $(k_{11},k_{12})$ and $(k_{21},k_{22})$ be the linearly independent eigenvectors related to eigenvalues $\lambda_1$ and $\lambda_2$, respectively, which yield
\begin{equation*}
\left(
\begin{array}{cc}
k_{i1} & k_{i2}\\
\end{array}
\right)
\left(
\begin{array}{cc}
-a-m_1-d_1\lambda_0+\mu-\lambda_i & b\\
a & -m_2-d_2\lambda_0+\mu-\lambda_i\\
\end{array}
\right)
=
\left(
\begin{array}{cc}
0 & 0\\
\end{array}
\right)
\end{equation*}
for $i=1,2$. Therefore
$$
(k_{11},k_{12})=(a,a+m_1+d_1\lambda_0-\mu+\lambda_1)=(a,a+m_1+d_1\lambda_0+c_1)
$$
and
$$
(k_{21},k_{22})=(m_2+d_2\lambda_0-\mu+\lambda_2,b)=(m_2+d_2\lambda_0+c_2,b).
$$

In the following, we consider the corresponding algebraic equations
\begin{equation*}
\left(
\begin{array}{cc}
a & a+m_1+d_1\lambda_0+c_1\\
m_2+d_2\lambda_0+c_2 & b\\
\end{array}
\right)
\left(
\begin{array}{c}
f_1(t)\\
f_2(t)\\
\end{array}
\right)
=
\left(
\begin{array}{c}
N_1e^{\lambda_1t}\\
N_2e^{\lambda_2t}\\
\end{array}
\right),
\end{equation*}
whence follows that
{\small $$
(f_1(t),f_2(t))=
\Big(\frac{N_1be^{\lambda_1t}-N_2(a+m_1+d_1\lambda_0+c_1)e^{\lambda_2t}}{C},\frac{N_1(a+m_1+d_1\lambda_0+c_1)e^{\lambda_1t}+N_2ae^{\lambda_2t}}{C}\Big)
$$}
for $t\in(0^+,\rho\tau]$, where $A_0:=a+m_1+d_1\lambda_0+c_1>0 \ \mbox{and }C:=ab-(a+m_1+d_1\lambda_0+c_1)(m_2+d_2\lambda_0+c_2)>0.$
In a similar way, for $((\rho\tau)^+,\tau]$, we have
$$
(f_3(t),f_4(t))=
\Big(\frac{N_3be^{\lambda_1t}-N_4A_0e^{\lambda_2t}}{C},\frac{N_3A_0e^{\lambda_1t}+N_4ae^{\lambda_2t}}{C}\Big).
$$

Since
$
f_3((\rho\tau)^+)=(1-\theta)f_1(\rho\tau)\,\,\mbox{and}\,\,f_4((\rho\tau)^+)=(1-\theta)f_2(\rho\tau),
$
we get
$
N_3=(1-\theta)N_1\,\,\mbox{and}\,\,N_4=(1-\theta)N_2.
$
Without loss of generality, we may assume that $N_1=1$. So,
$$
(f_1(t),f_2(t))=
\Big(\frac{be^{\lambda_1t}-N_2A_0e^{\lambda_2t}}{C},\frac{A_0e^{\lambda_1t}+N_2ae^{\lambda_2t}}{C}\Big).
$$

Recalling
$
f_1(0^+)=(1+\delta)f_3(\tau)\,\,\mbox{and}\,\,f_2(0^+)=f_4(\tau),
$
we obtain
\begin{eqnarray}
\left\{
\begin{aligned}
&b-A_0N_2=(1+\delta)[N_3be^{\lambda_1\tau}-N_4A_0e^{\lambda_2\tau}], \\[1mm]
&(a+m_1+d_1\lambda_0+c_1)+aN_2=N_3A_0e^{\lambda_1\tau}+N_4ae^{\lambda_2\tau}.
\end{aligned} \right.
\label{6c}
\end{eqnarray}
For abbreviation, we further denote in \eqref{6c} that
$$B_{11}=b, \,B_{12}=(1+\delta)(1-\theta)b,\,B_{13}=(1+\delta)(1-\theta)A_0e^{(c_2-c_1)\tau},$$
$$\Lambda=e^{\lambda_1\tau},\,B_{21}=a,\,B_{22}=(1-\theta)A_0,\,B_{23}=(1-\theta)ae^{(c_2-c_1)\tau},$$
then leads us to that
\begin{equation}
\Lambda=\frac{B_{11}-A_0N_2}{B_{12}-B_{13}N_2},\,\,\Lambda=\frac{A_0+B_{21}N_2}{B_{22}+B_{23}N_2}.
\label{7c}
\end{equation}

The explicit solution $(N_2,\Lambda)$ of the above system is not easy to be given, so in the following we will develop the existence of positive eigenfunction by image.
\begin{figure}[!ht]
\centering
\subfigure[]{ {
\includegraphics[width=0.29\textwidth]{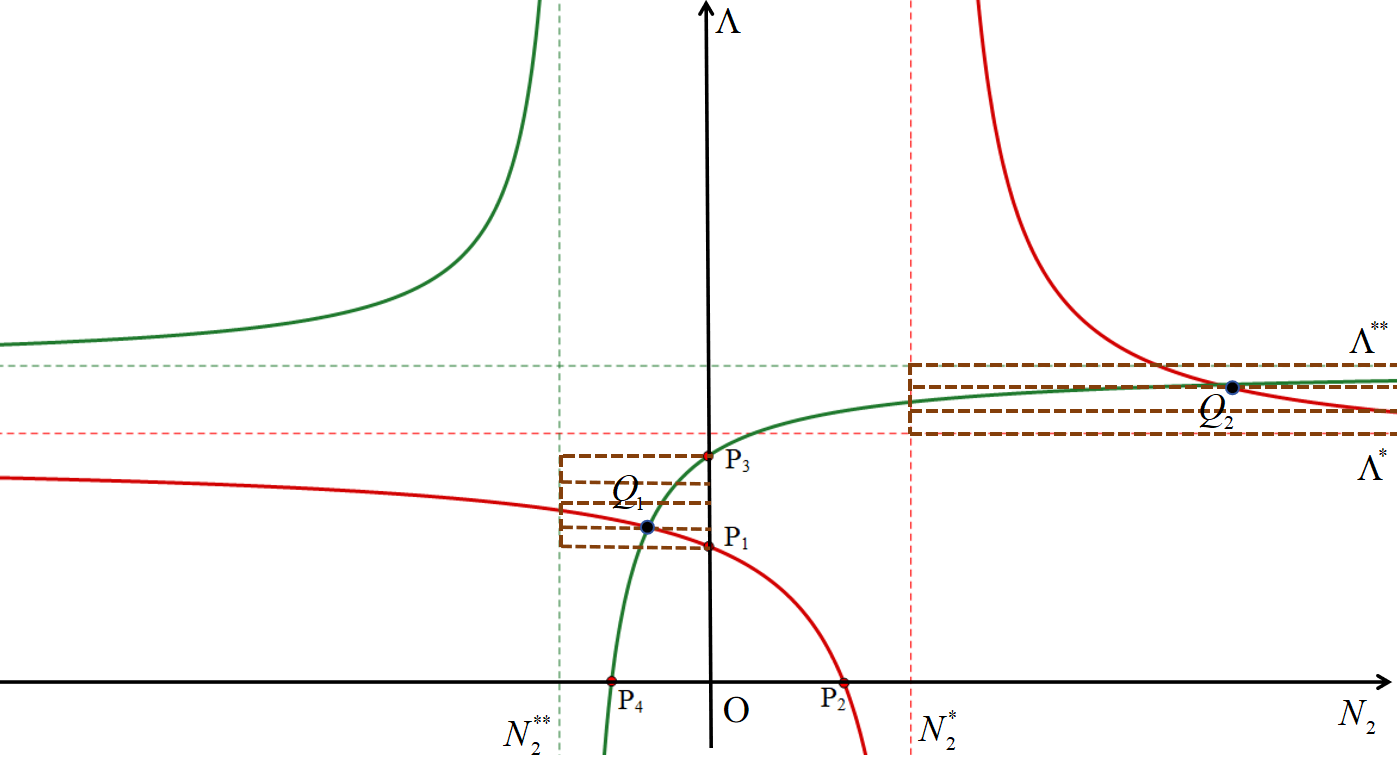}
} }
\subfigure[]{ {
\includegraphics[width=0.29\textwidth]{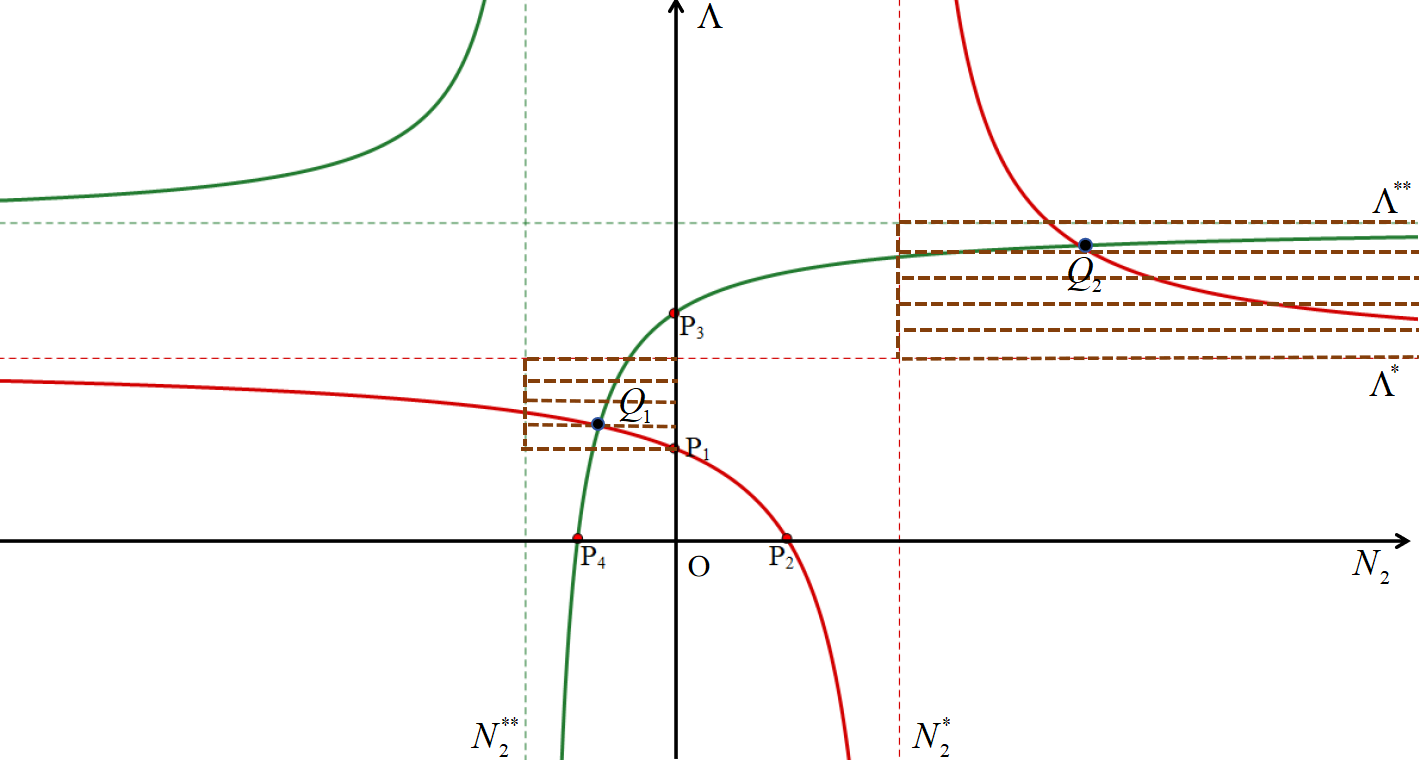}
} }
\subfigure[]{ {
\includegraphics[width=0.29\textwidth]{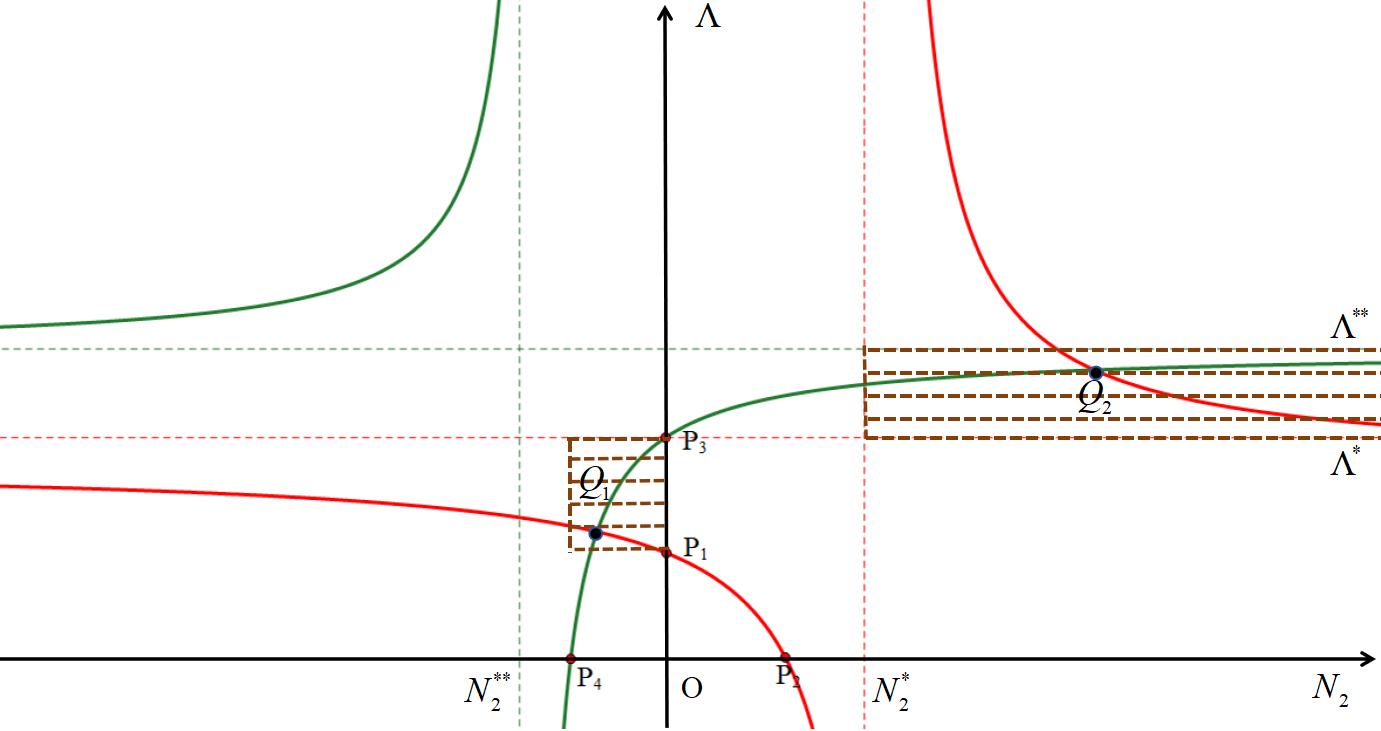}
} }
\caption{\scriptsize The solution curve of $(N_2,\Lambda)$ in \eqref{7c}. The red line represents the solution curve of equation $\Lambda=\frac{B_{11}-A_0N_2}{B_{12}-B_{13}N_2}$, which is strictly decreasing with respect to $N_2$ in interval $(-\infty,\frac{b}{A_0e^{(c_2-c_1)\tau}})\bigcup(\frac{b}{A_0e^{(c_2-c_1)\tau}},\infty)$. Also, it goes through fixed points $P_1(0,\frac{1}{(1+\delta)(1-\theta)})$ and $P_2(\frac{b}{A_0},0)$. One easily checks that $N_2^*=\frac{b}{A_0e^{(c_2-c_1)\tau}}$ and $\Lambda^*=\frac{1}{(1+\delta)(1-\theta)e^{(c_2-c_1)\tau}}$ are two asymptotics. Similarly, the green line is the solution curve of equation $\Lambda=\frac{A_0+B_{21}N_2}{B_{22}+B_{23}N_2}$, which is strictly increasing with respect to $N_2$ in interval $(-\infty,\frac{A_0}{ae^{(c_2-c_1)\tau}})\bigcup(\frac{A_0}{ae^{(c_2-c_1)\tau}},\infty)$. It goes through the fixed points $P_3(0,\frac{1}{1-\theta})$ and $P_4(-\frac{A_0}{a},0)$, and $N_2^{**}=-\frac{A_0}{ae^{(c_2-c_1)\tau}}$, $\Lambda^{**}=\frac{1}{(1-\theta)e^{(c_2-c_1)\tau}}$ are two asymptotics. Graph (a) is the case of $(1+\delta)e^{(c_2-c_1)\tau}<1$, Graph (b) is the case of $(1+\delta)e^{(c_2-c_1)\tau}>1$ and Graph (c) is the case of $(1+\delta)e^{(c_2-c_1)\tau}=1$, respectively.
}
\label{tu4}
\end{figure}

If $(1+\delta)e^{(c_2-c_1)\tau}<1$, it can be seen from Fig. \ref{tu4}(a) that these two lines have two intersections $Q_1(N_2,\Lambda)\in D_1:=(-\frac{A_0}{ae^{(c_2-c_1)\tau}},0)\times(\frac{1}{(1+\delta)(1-\theta)},\frac{1}{1-\theta})$ and $Q_2(N_2,\Lambda)\in D_2:=(\frac{b}{A_0e^{(c_2-c_1)\tau}},\infty)\times(\frac{1}{(1+\delta)(1-\theta)e^{(c_2-c_1)\tau}},\frac{1}{(1-\theta)e^{(c_2-c_1)\tau}})$. Since $f_1(t)f_2(t)<0$ in $D_2$, problem \eqref{7c} admits a unique solution $Q_1(N_2,\Lambda)\in D_1$ with $f_1(t)f_2(t)>0$, which indicates that constant-coefficient problem \eqref{a02} admits a unique solution $(\mu,\phi,\psi)$ with $\phi,\psi>0$. See also $Q_1(N_2,\Lambda)\in D_3:=(-\frac{A_0}{ae^{(c_2-c_1)\tau}},0)\times(\frac{1}{(1+\delta)(1-\theta)},\frac{1}{(1+\delta)(1-\theta)e^{(c_2-c_1)\tau}})$ in Fig. \ref{tu4}(b) with $(1+\delta)e^{(c_2-c_1)\tau}>1$ and $Q_1(N_2,\Lambda)\in D_5:=(-\frac{A_0}{a},0)\times(\frac{1}{(1+\delta)(1-\theta)},\frac{1}{1-\theta})$ in Fig. \ref{tu4}(c) with $(1+\delta)e^{(c_2-c_1)\tau}=1$.

Therefore, if all coefficients are constant in eigenvalue problem \eqref{a02}, it follows from the mathematical analysis and image method that the principal eigen-pair $(\mu,\phi,\psi)$ exists with $\phi(t,x),\psi(t,x)>0$ in $t\in(0^+,\tau]$ and $x\in\overline \Omega$.

\vspace{1mm}
In the sequel, we give an example to show the expressions of eigen-pair in case of dual effects of birth and harvesting perturbations. For such an one-single logistic model
\begin{eqnarray}
\left\{
\begin{aligned}
&\phi_t-d\phi_{xx}=m(t)\phi+\mu_1 \phi, && t\in(0^+,\rho\tau]\bigcup ((\rho\tau)^+,\tau],\,x \in \Omega,  \\
&\phi(t,x)=0, && t\in [0, \tau],\,\,x \in \partial \Omega,\\
&\phi(0,x)=\phi(\tau,x), && x\in \overline \Omega,\\
&\phi(0^+, x)=(1+\delta)\phi(0,x),&& x\in \overline \Omega,\\
&\phi((\rho\tau)^+, x)=(1-\theta)\phi(\rho\tau, x),&&x\in \overline \Omega,
\end{aligned} \right.
\label{c}
\end{eqnarray}
we now present the principal eigenvalue and positive eigenfunction by direct calculations.
\begin{exm}\label{aaa1}
For one-single logistic model \eqref{c}, the principal eigenvalue of the problem \eqref{a01} can be precisely expressed as
\begin{equation}\label{abc}
\mu_1=-\frac{\ln[(1+\delta)(1-\theta)]}{\tau}+d\lambda_1-\frac{\int_{0}^{\tau}m(t)dt}{\tau}
\end{equation}
with positive eigenfunction $\phi(t,x)(=f(t)\psi(x))$ in $(t,x)\in [0,\tau]\times\overline\Omega$, where $\lambda_1(>0)$ is the principal eigenvalue of $-\Delta$ in $\Omega$ with homogeneous Dirichlet boundary condition, and the positive function $f(t)$ is defined by
\begin{eqnarray*}
f(t)=
\begin{cases}
\displaystyle 1, &t=0, \\[1mm]
\displaystyle (1+\delta), &t=0^+,\\[1mm]
\displaystyle (1+\delta)e^{\int_{0}^{t}m(t)dt+(\mu_1-d\lambda_1)t}, &t\in(0^+,\rho\tau],\\[1mm]
\displaystyle (1-\theta)(1+\delta)e^{\int_{0}^{\rho\tau}m(t)dt+(\mu_1-d\lambda_1)\rho\tau}, &t=(\rho\tau)^+,\\[1mm]
\displaystyle (1-\theta)(1+\delta)e^{\int_{0}^{t}m(t)dt+(\mu_1-d\lambda_1)t},&t\in ((\rho\tau)^+,\tau].
\end{cases}
\end{eqnarray*}
\end{exm}
\bpf Substituting $\phi(t,x)=f(t)\psi(x)$ into equations in \eqref{c} and proceeding similarly as done in \cite[Theorem 2.1]{M}, one easily checks that \eqref{abc} holds.
In fact, for $t\in(0^+,\rho\tau]$, it is clear that
$\int_{0^+}^{t}{f'(t)}/{f(t)}dt=\int_{0}^{t}m(t)dt+(\mu_1-d\lambda_1)t$
so that
\begin{equation}\label{a055}
f(t)=f(0^+)e^{\int_{0}^{t}m(t)dt+(\mu_1-d\lambda_1)t}=(1+\delta)f(0)e^{\int_{0}^{t}m(t)dt+(\mu_1-d\lambda_1)t}.
\end{equation}
Similarly, for $t\in((\rho\tau)^+,\tau]$, we have $\int_{(\rho\tau)^+}^{t}f'(t)/f(t)dt=\int_{\rho\tau}^{t}m(t)dt+(\mu_1-d\lambda_1)(t-\rho\tau),$
that  leads to
\begin{eqnarray}
\begin{array}{ll}
f(t)&=f((\rho\tau)^+)e^{\int_{\rho\tau}^{t}m(t)dt+(\mu_1-d\lambda_1)(t-\rho\tau)}\\[1mm]
&=(1-\theta)f(\rho\tau)e^{\int_{\rho\tau}^{t}m(t)dt+(\mu_1-d\lambda_1)(t-\rho\tau)}\\[1mm]
&=(1-\theta)(1+\delta)f(0)e^{\int_{0}^{t}m(t)dt+(\mu_1-d\lambda_1)t}.
\label{a057}
\end{array}
\end{eqnarray}

 Next, let $f(0)=1$, then $f(0^+)=(1+\delta)f(0)=(1+\delta)$. From \eqref{a055}, we have $f((\rho\tau)^+)=(1-\theta)f(\rho\tau)=(1-\theta)(1+\delta)e^{\int_{0}^{\rho\tau}m(t)dt+(\mu_1-d\lambda_1)\rho\tau}$. So, it follows from \eqref{a057} and \eqref{abc} that $f(\tau)=f(0)$ due to the fact that
 $f(\tau)=(1-\theta)(1+\delta)f(0)e^{\int_{0}^{\tau}m(t)dt+(\mu_1-d\lambda_1)\tau}=1.$
\epf

\vspace{1mm}
Before showing the long-time behavior of the solution of the problem \eqref{a01}, we first investigate the properties of the principal eigenvalue of \eqref{a02}. To do this, let us consider the following auxiliary problem
\begin{eqnarray}
\left\{
\begin{aligned}
&-\phi^*_{t}=d_1\phi^*_{xx}+{a(t)}\psi^*-(a(t)+m_1(t))\phi^*+\lambda\phi^*, &&t\in(0^+,\rho\tau]\cup ((\rho\tau)^+,\tau],\,\,x \in \Omega,\\
&-\psi^*_{t}=d_2\psi^*_{xx}+{b(t)}\phi^*-m_2(t)\psi^*+\lambda\psi^*, &&t\in(0^+,\rho\tau]\cup ((\rho\tau)^+,\tau],\,\,x \in \Omega,  \\
&\phi^*(0,x)=\phi^*(\tau, x), \,\psi^*(0,x)=\psi^*(\tau, x), && x\in \overline \Omega,\\
&\phi^*(0^+, x)=1/(1+\delta)\phi^*(0,x),&& x\in \overline \Omega,\\
&\phi^*((\rho\tau)^+, x)=1/(1-\theta)\phi^*(\rho\tau,x),&& x\in \overline \Omega,\\
&\psi^*((\rho\tau)^+, x)=1/(1-\theta)\psi^*(\rho\tau,x),&& x\in \overline \Omega,\\
&\phi^*(t,x)=\psi^*(t,x)=0, && t\in [0,\tau],\,\,x \in \partial \Omega.\\
\end{aligned} \right.
\label{a03}
\end{eqnarray}

\begin{lem}\label{le1}
One has that $\lambda=\mu$, where $\lambda$ and $\mu$ are the principal eigenvalues of \eqref{a03} and \eqref{a02}, respectively, and $(\phi^*,\psi^*)$ is the first eigenfunction associate to $\lambda$.
\end{lem}
\bpf
Multiplying the first equation in \eqref{a02} by $\phi^*$ and the first equation in \eqref{a03} by $\phi$, respectively, and then subtracting  one from another, we obtain
\begin{align}
\phi_t\phi^*+\phi_t^*\phi-d_1( \phi_{xx}\phi^*- \phi^*_{xx}\phi)=b(t)\phi^*\psi-a(t)\psi^*\phi(\mu-\lambda)\phi\phi^*.
\label{3.25}
\end{align}
Recalling the pulse and periodic conditions in \eqref{a02} and \eqref{a03}, we get
$\int_{\Omega}( \phi_{xx}\phi^*- \phi^*_{xx}\phi)dx=0,$
whence follows that
\begin{align*}
&\Big(\int_{0^+}^{\rho\tau}+\int_{(\rho\tau^)+}^{\tau}\Big)(\phi_t\phi^*+\phi_t^*\phi)dt\\
=&\phi(\rho\tau,x)\phi^*(\rho\tau,x)-\phi(0^+,x)\phi^*(0^+,x)+\phi(\tau,x)\phi^*(\tau,x)-\phi((\rho\tau)^+,x)\phi^*((\rho\tau)^+,x)\\
=&0.
\end{align*}

So by integrating both sides of the equations in \eqref{3.25} over $t\in(0^+,\rho\tau]\cup((\rho\tau)^+,\tau]$ and $x\in\Omega$, yields
\begin{equation}\label{314}
    \Big(\int_{0^+}^{\rho\tau}+\int_{(\rho\tau)^+}^{\tau}\Big)\int_{\Omega}[b(t)\phi^*\psi-a(t)\psi^*\phi+(\mu-\lambda)\phi\phi^*]dxdt=0.
\end{equation}

Similarly, we can multiply the second equations in \eqref{a02} by $\psi^*$ and in \eqref{a03} by $\psi$, to obtain, after
integrating over $t\in(0^+,\rho\tau]\cup((\rho\tau)^+,\tau]$ and $x\in\Omega$, that
\begin{equation}\label{315}
  \Big(\int_{0^+}^{\rho\tau}+\int_{(\rho\tau)^+}^{\tau}\Big)\int_{\Omega}[a(t)\psi^*\phi-b(t)\phi^*\psi+(\mu-\lambda)\psi\psi^*]dxdt=0
\end{equation}
holds.
Therefore, it follows from \eqref{314} and \eqref{315}, that
$$
(\mu-\lambda)\Big(\int_{0^+}^{\rho\tau}+\int_{(\rho\tau)^+}^{\tau}\Big)\int_{\Omega}(\phi\phi^*+\psi\psi^*)dxdt=0
$$
holds, whence follows that $\mu=\lambda$ since $\Big(\int_{0^+}^{\rho\tau}+\int_{(\rho\tau)^+}^{\tau}\Big)\int_{\Omega}(\phi\phi^*+\psi\psi^*)dxdt>0.$
This ends the proof.
\epf

\vspace{1mm}
To emphasize the impact of coefficients about birth pulse $\delta$, harvesting pulse $\theta$ and the domain $\Omega$, we denote $\mu:=\mu(\delta, \theta, \Omega).$
\begin{thm}
The following statements hold:
\begin{enumerate}
    \item[$(i)$] $\mu(\delta, \theta, \Omega)$ is strictly monotonic decreasing with respect to $\delta$ for any given $\theta$ and $\Omega$;
      \item[$(ii)$] $\mu(\delta, \theta, \Omega)$ is strictly monotonic increasing with respect to $\theta$ for any given $\delta$ and $\Omega$;
        \item[$(iii)$] $\mu(\delta, \theta, \Omega)$ is strictly monotonic decreasing with respect to $\Omega$ for any given $\delta$ and $\theta$.
\end{enumerate}
\label{5c01}
\end{thm}
\bpf
Item $(i)$. Since $\phi$, $\psi$ and $\mu$ are smooth functions on $\delta$ and $\theta$ by perturbation results in \cite{K}, we obtain by differentiating both sides of the equations in problem \eqref{a02}, with respect to $\delta$, that
\begin{eqnarray}
\left\{
\begin{aligned}
&\phi'_{t}=d_1\phi'_{xx}+{b(t)}\psi'-(a(t)+m_1(t))\phi'+\mu'\phi+\mu\phi', &&t\in(0^+,\rho\tau]\cup ((\rho\tau)^+,\tau],\,\,x \in \Omega,  \\
&\psi'_{t}=d_2\psi'_{xx}+{a(t)}\phi'-m_2(t)\psi'+\mu'\psi+\mu\psi', &&t\in(0^+,\rho\tau]\cup ((\rho\tau)^+,\tau],\,\,x \in \Omega,  \\
&\phi'(0,x)=\phi'(\tau, x), \,\psi'(0,x)=\psi'(\tau, x), && x\in \overline \Omega,\\
&\phi'(0^+, x)=\phi(0,x)+(1+\delta)\phi'(0,x),&& x\in \overline \Omega,\\
&\phi'((\rho\tau)^+, x)=(1-\theta)\phi'(\rho\tau,x),&& x\in \overline \Omega,\\
&\psi'((\rho\tau)^+, x)=(1-\theta)\psi'(\rho\tau,x),&& x\in \overline \Omega,\\
&\phi'(t,x)=\psi'(t,x)=0, && t\in [0,\tau],\,\,x \in \partial \Omega\\
\end{aligned} \right.
\label{3.26}
\end{eqnarray}
hold. Similarly, as done in the proof of Lemma \ref{le1}, by multiplying the first equation in \eqref{3.26} by $\phi^*$ and the first equation in \eqref{a03} by $\phi'$, then subtracting one from another, and integrating the resultant equation over $t\in(0^+,\rho\tau]\cup((\rho\tau)^+,\tau]$ and $x\in\Omega$, we obtain
$$
\frac{-1}{1+\delta}\int_{\Omega}\phi(0,x)\phi^*(0,x)dx
=\Big(\int_{0^+}^{\rho\tau}+\int_{(\rho\tau)^+}^{\tau}\Big)\int_{\Omega}[b(t)\phi^*\psi'-a(t)\psi^*\phi'+\mu'\phi\phi^*]dxdt,
$$
and, agian in a similar way as done in the proof of  Lemma \ref{le1}, we have
$$
\Big(\int_{0^+}^{\rho\tau}+\int_{(\rho\tau)^+}^{\tau}\Big)\int_{\Omega}[a(t)\psi^*\phi'-b(t)\phi^*\psi'+\mu'\psi\psi^*]dxdt=0.
$$

Thus, adding these two equations to obtain
$$
\mu'=\frac{\frac{-1}{1+\delta}\int_{\Omega}\phi(0,x)\phi^*(0,x)dx}
{(\int_{0^+}^{\rho\tau}+\int_{(\rho\tau)^+}^{\tau})\int_{\Omega}(\phi\phi^*+\psi\psi^*)dxdt}<0.
$$

\noindent Item $(ii)$. By differentiating the both sides of equations in problem \eqref{a02} with respect to $\beta$, and then using the same strategy as done in $(i)$, we obtain
$$
\mu'=\frac{\frac{1}{1-\theta}\int_{\Omega}[\phi(\rho\tau,x)\phi^*(\rho\tau,x)+\psi(\rho\tau,x)\psi^*(\rho\tau,x)]dx}{(\int_{0^+}^{\rho\tau}
+\int_{(\rho\tau)^+}^{\tau})\int_{\Omega}(\phi\phi^*+\psi\psi^*)dxdt}>0.
$$

\noindent Item $(iii)$. Suppose $\Omega_1\subseteq \Omega_2\subseteq R^N$ and $\Omega_2 \backslash \Omega_1$ is nonempty. Let $(\phi,\psi,\mu(\Omega_2))$ and $(\phi^*,\psi^*,\mu(\Omega_1))$ be the principal eigen-pairs to problems \eqref{a02} in $\Omega_2$ and \eqref{a03} in $\Omega_1$,  respectively. That is,
\begin{eqnarray}
\left\{
\begin{aligned}
&\phi_{t}=d_1\phi_{xx}+{b(t)}\psi-(a(t)+m_1(t))\phi+\mu(\Omega_2)\phi, &&t\in(0^+,\rho\tau]\cup ((\rho\tau)^+,\tau],\,\,x \in \Omega_2,  \\
&\psi_{t}=d_2\psi_{xx}+{a(t)}\phi-m_2(t)\psi+\mu(\Omega_2)\psi, &&t\in(0^+,\rho\tau]\cup ((\rho\tau)^+,\tau],\,\,x \in \Omega_2,  \\
&\phi(0,x)=\phi(\tau, x), \,\psi(0,x)=\psi(\tau, x), && x\in \overline \Omega_2,\\
&\phi(0^+, x)=(1+\delta)\phi(0,x),&& x\in \overline \Omega_2,\\
&\phi((\rho\tau)^+, x)=(1-\theta)\phi(\rho\tau,x),&& x\in \overline \Omega_2,\\
&\psi((\rho\tau)^+, x)=(1-\theta)\psi(\rho\tau,x),&& x\in \overline \Omega_2,\\
&\phi(t,x)=\psi(t,x)=0, && t\in [0,\tau],\,\,x \in \partial \Omega_2,
\end{aligned} \right.
\label{3.8}
\end{eqnarray}
and
\begin{eqnarray}
\left\{
\begin{aligned}
&-\phi^*_{t}=d_1\phi^*_{xx}+{a(t)}\psi^*-(a(t)+m_1(t))\phi^*+\mu(\Omega_1)\phi^*, &&t\in(0^+,\rho\tau]\cup ((\rho\tau)^+,\tau],\,\,x \in \Omega_1,\\
&-\psi^*_{t}=d_2\psi^*_{xx}+{b(t)}\phi^*-m_2(t)\psi^*+\mu(\Omega_1)\psi^*, &&t\in(0^+,\rho\tau]\cup ((\rho\tau)^+,\tau],\,\,x \in \Omega_1,  \\
&\phi^*(0,x)=\phi^*(\tau, x), \,\psi^*(0,x)=\psi^*(\tau, x), && x\in \overline \Omega_1,\\
&\phi^*(0^+, x)=1/(1+\delta)\phi^*(0,x),&& x\in \overline \Omega_1,\\
&\phi^*((\rho\tau)^+, x)=1/(1-\theta)\phi^*(\rho\tau,x),&& x\in \overline \Omega_1,\\
&\psi^*((\rho\tau)^+, x)=1/(1-\theta)\psi^*(\rho\tau,x),&& x\in \overline \Omega_1,\\
&\phi^*(t,x)=\psi^*(t,x)=0, && t\in [0,\tau],\,\,x \in \partial \Omega_1,
\end{aligned} \right.
\label{3.9}
\end{eqnarray}
hold. So, by multiplying the first equation in \eqref{3.8} by $\phi^*$ and the first one in \eqref{3.9} by $\phi$, then subtracting one from another, and integrating the resultant equation over $t\in(0^+,\rho\tau]\bigcup((\rho\tau)^+,\tau]$ and $x\in\Omega_1$, we assert that
\begin{align*}
&d_1\Big(\int_{0^+}^{\rho\tau}+\int_{(\rho\tau)^+}^{\tau}\Big)(\phi_{x}^*\phi)|_{x\in\partial \Omega_1}dt\\
=&\Big(\int_{0^+}^{\rho\tau}+\int_{(\rho\tau)^+}^{\tau}\Big)\int_{\Omega}
[b(t)\psi\phi^*-a(t)\phi\psi^*+(\mu(\Omega_2)-\mu(\Omega_1))\phi\phi^*]dxdt.
\end{align*}

Following similar reasoning, we have
\begin{align*}
&d_2\Big(\int_{0^+}^{\rho\tau}+\int_{(\rho\tau)^+}^{\tau}\Big)(\psi_{x}^*\psi)|_{x\in\partial \Omega_1}dt\\
=&\Big(\int_{0^+}^{\rho\tau}+\int_{(\rho\tau)^+}^{\tau}\Big)\int_{\Omega}
[a(t)\phi\psi^*-b(t)\psi\phi^*+(\mu(\Omega_2)-\mu(\Omega_1))\psi\psi^*]dxdt.
\end{align*}

So, by adding these two equations, we get
$$
\mu(\Omega_2)-\mu(\Omega_1)=\frac{(\int_{0^+}^{\rho\tau}+\int_{(\rho\tau)^+}^{\tau})[(d_1\phi_{x}^*\phi)|_{x\in\partial \Omega_1}+(d_2\psi_{x}^*\psi)|_{x\in\partial \Omega_1}]dt}{(\int_{0^+}^{\rho\tau}+\int_{(\rho\tau)^+}^{\tau})\int_{\Omega}(\phi\phi^*+\psi\psi^*)dxdt},
$$
whence follows that $\mu(\Omega_2)-\mu(\Omega_1)<0$ since $\phi,\psi\geq,\not\equiv 0$ and $\phi_{x}^*,\psi_{x}^*<0$ in $\partial \Omega_1$ by the strong maximum principle. This ends the proof.
\epf

\section{The dynamical behavior of the solution}
In this section, the principal eigenvalue $\mu:=\mu(\delta, \theta, \Omega)$, defined in Section 3, and the method of upper and lower solutions will be employed to study the asymptotic behavior of the solution of problem \eqref{a01}.
\begin{thm}\label{ab}
 If $\mu(\delta, \theta, \Omega)>0$, then the solution $(u_1(t,x),u_2(t,x))$ to problem \eqref{a01} satisfies $\lim\limits_{t\to\infty}{(u_1(t,x),u_2(t,x))}=(0,0)$ uniformly for $x\in \overline \Omega$.
\end{thm}
\bpf
Let
$$
\bar{u}_1(t,y)=Ke^{-\mu t}\phi(t,x)\,\,\, and\,\,\, \bar{u}_2(t,x)=Ke^{-\mu t}\psi(t,x),
$$
where $0<\parallel \phi\parallel+\parallel\psi\parallel\leq 1$ and $K$ is a real positive constant such that
$K\phi(0,x)\geq u_{1,0}(x)\ \mbox{and } K\psi(0,x)\geq u_{2,0}(x).$

For $t\in((n\tau)^+,(n+\rho)\tau]\cup (((n+\rho)\tau)^+,(n+1)\tau]$ and $x \in \Omega$, it follows from \eqref{a02} that
$$\begin{array}{ll}
&\bar{u}_{1t}-d_1\bar{u}_{1xx}-b(t)\bar{u}_2+(a(t)+m_1(t))\bar{u}_1+\alpha_1(t)\bar{u}_1^2 \\[1mm]
&\geq Ke^{-\mu t}[-\mu\phi+\phi_t-d_1\phi_{xx}-b(t)\psi+(a(t)+m_1(t))\phi]\\[1mm]
&=0.
\end{array}$$
In a similar way, we have
$\bar{u}_{2t}-d_12\bar{u}_{2xx}-a(t)\bar{u}_2+m_2(t)\bar{u}_2+\alpha_2(t)\bar{u}_2^2\geq0.$

Besides, the impulsive conditions, for $x\in\overline \Omega$, satisfies
$$\bar{u}_1((n\tau)^+,x)=(1+\delta)\bar{u}_1(n\tau,x),\,\,\bar{u}_i(((n+\rho)\tau)^+,x)=(1-\theta)\bar{u}_i((n+\rho)\tau,x),\,i=1,2,$$
and the initial value $(\bar u_1,\bar u_2)(0,x)\geq (u_{1,0},u_{2,0})(x)$ for $x\in \overline \Omega$.

Therefore, a comparison principle asserts
$(0,0) \leq  (u_1,u_2)(t,x)\leq(\bar{u}_1,\bar u_2)(t,x)\ \mbox{for }x\in\overline \Omega\ \mbox{and }t>0,$
whence follows together with $\lim\limits_{t\to\infty}{(\bar u_1, \bar u_2)(t,x)}=(0,0)$, due to $\mu>0$,  that $\lim\limits_{t\to\infty}{(u_1,u_2)(t,x)}=(0,0)\ \mbox{uniformly for }  x\in\overline \Omega.$
This ends the proof.
\epf

\vspace{1mm}
Below, let us research the long-time behavior of the solution to problem \eqref{a01} with pulses in case  $\mu(\delta, \theta, \Omega)<0$. To do this, we first introduce the  corresponding periodic problem
\begin{eqnarray}
\left\{
\begin{array}{lll}
U_{1t}=d_1U_{1xx}+{b(t)}U_2-(a(t)+m_1(t))U_1-\alpha_1(t)U_1^2,  &(t,x)\in \Omega_{\rho,\tau}^0,\\[1mm]
U_{2t}=d_2U_{2xx}+{a(t)}U_1-m_2(t)U_2-\alpha_2(t)U_2^2, & (t,x)\in \Omega_{\rho,\tau}^0,\\[1mm]
U_1(0^+,x)=(1+\delta)u_1(0,x), & x\in \overline \Omega, \\[1mm]
U_1((\rho\tau)^+,x)=(1-\theta)U_1(\rho\tau,x), & x\in \overline \Omega, \\[1mm]
U_2((\rho\tau)^+,x)=(1-\theta)U_2(\rho\tau,x), & x\in \overline \Omega, \\[1mm]
U_1(t,x)=U_2(t,x)=0, & t\in(0,\tau],\,\,x \in \partial \Omega, \\[1mm]
U_1(0,x)=U_1(\tau,x), U_2(0,x)=U_2(\tau,x),&x\in \overline \Omega,
\end{array} \right.
\label{c01}
\end{eqnarray}
where $\Omega_{\rho,\tau}^0$ is defined by $\Omega_{\rho,\tau}^n$  as $n=0$.
\begin{thm} \label{ad}
Assume that $\mu(\delta, \theta, \Omega)<0$. Then:
\begin{enumerate}
    \item[$(i)$] the periodic problem \eqref{c01} admits a unique solution $(U_1^*(t,x),U_2^*(t,x))$;
    \item[$(ii)$] the unique solution $(u_1(t,x),u_2(t,x))$ of the problem \eqref{a01} satisfies
    $$\lim_{m\to \infty} (u_1,u_2)(t+m\tau,x)=(U_1^*,U_2^*)(t,x)\ \mbox{for any } t\geq 0\ \mbox{and uniformly for }x\in \overline \Omega,$$
for each $(u_{1,0},u_{2,0})\in [C(\overline \Omega)]^2$ such that $(u_{1,0},u_{2,0})\geq (0,0)$ and $(u_{1,0},u_{2,0})\neq (0,0)$, where $(U_1^*,U_2^*)$ is the solution of problem \eqref{c01} given by $(i)$ just above.
\end{enumerate}
\end{thm}
\bpf
Item $(i)$. It is clear by the equations in \eqref{a02} that $\mu:=\mu(\delta, \theta, \Omega):=\mu(b(t), a(t), \delta, \theta, \Omega)$
is nonincreasing in $b(t)$ and $a(t)$ so that $\mu(b^M,a^M,\delta,\theta,\Omega)<0$, and there exists a positive constant $K^*$ such that
\begin{eqnarray*}
\left\{
\begin{aligned}
&\phi_{t}=d_1\phi_{xx}+({b^M}-K^*)\psi-(a^m+m_1^m)\phi+\mu^\triangle\phi,\\
&\psi_{t}=d_2\psi_{xx}+({a^M}-K^*)\phi-m_2^m\psi+\mu^\triangle\psi\\
\end{aligned} \right.
\end{eqnarray*}
with $\mu^\triangle:=\mu^\triangle(b^M-K^*,a^M-K^*,\delta,\theta,\Omega)>0$. Next, define
$$\Omega_\varepsilon:=\{x\in\Omega:\,\textrm{dist}\,(x,\partial \Omega)>\varepsilon\},$$
and denote by $(\mu_{\varepsilon}^\triangle,\Phi,\Psi)$ its associate eigen-pair with $\max_{(t,x)\in [0,\tau]\times \overline \Omega}\{\Phi(t,x),\Psi(t,x)\}= 1$, that is, one has
\begin{eqnarray*}
\left\{
\begin{aligned}
&\Phi_{t}=d_1\Phi_{xx}+({b^M}-K^*\chi_{\Omega_\varepsilon})\Psi-(a^m+m_1^m)\Phi+\mu_\varepsilon^\triangle\Phi,\\
&\Psi_{t}=d_2\Psi_{xx}+({a^M}-K^*\chi_{\Omega_\varepsilon})\Phi-m_2^m\Psi+\mu_\varepsilon^\triangle\Psi,\\
\end{aligned} \right.
\end{eqnarray*}
where $\chi_{\Omega_\varepsilon}(x)=1\, \textrm{for}\, x\in \Omega_\varepsilon\, \textrm{and}\, \chi_{\Omega_\varepsilon}(x)=0\, \textrm{for}\, x\in \Omega/ \Omega_\varepsilon$. Since $\mu^\triangle>0$, we can choose a sufficiently small $\varepsilon>0$ such that $\mu_{\varepsilon}^\triangle:=\mu_{\varepsilon}^\triangle( a^M-K\chi_{\Omega_\varepsilon},b^M-K\chi_{\Omega_\varepsilon},\alpha,\beta)>0.$
So, by defining  $$(\overline{U}_1,\overline{U}_2)=(M\Phi,M\Psi),$$ where $M$ is a real parameter, we obtain that
\begin{align*}
&\overline{U}_{1t}-d_1 \overline{U}_{1xx}-b(t)\overline U_2+(a(t)+m_1(t))\overline U_1+\alpha_1 U_1^2\\
=&M[\Phi_t-d_1\Phi_{xx}-b(t)\Psi+(a(t)+m_1(t))\Phi+\alpha_1M\Phi^2]\\
\geq&M(-K^*\chi_{\Omega_\varepsilon}\Psi+\mu_{\varepsilon}^\triangle\Phi+\alpha_1M\Phi^2)\\
>&M(-K^*\chi_{\Omega_\varepsilon}\Psi+\alpha_1M\Phi^2)\\
\geq&0,
\end{align*}
and
\begin{align*}
&\overline{U}_{2t}-d_2 \overline{U}_{2xx}-a(t)\overline U_1+m_2(t)\overline U_2+\alpha_2(t) U_2^2\\
=&M[\Psi_t-d_2\Psi_{xx}-a(t)\Phi+m_2(t)\Psi+\alpha_2M\Psi^2]\\
\geq&M(-K^*\chi_{\Omega_\varepsilon}\Phi+\mu_{\varepsilon}^\triangle\Psi+\alpha_2M\Psi^2)\\
\geq&0
\end{align*}
hold for an $M>1$ large enough. Besides, it is immediately to verify that
$$\overline {U}_1(0^+,x)=(1+\delta)\overline{U}_1(0,x)\ \mbox{and }\overline {U}_i((\rho\tau)^+,x)=(1-\theta)\overline{U}_i(\rho\tau,x)\ \mbox{for }i=1,2,$$
whence follows that  $(\overline U_1,\overline U_2)$ is an upper solution to problem \eqref{c01}.

Below, let us build a lower solution to problem \eqref{c01}. Define
\begin{eqnarray*}
\underline{U}_1(t,x)=
\begin{cases}
\displaystyle \varepsilon{\phi}(0,x), &t=0,\,\,x\in\overline\Omega, \\[3mm]
\displaystyle \varepsilon{\frac{\kappa_1}{1+\delta}}\phi(0^+,x), &t=0^+,\,\,x\in\overline\Omega,\\[3mm]
\displaystyle \varepsilon{\frac{\kappa_1}{1-\theta}}\phi((\rho\tau)^+,x), &t=(\rho\tau)^+,\,\,x\in\overline\Omega,\\[3mm]
\displaystyle \varepsilon{\frac{\kappa_1}{1-\theta}}e^{-\frac{\mu}{2}t}\phi(t,x),&t\in(0^+,\rho\tau]\cup((\rho\tau)^+,\tau],\,\,x\in\overline\Omega,
\end{cases}
\end{eqnarray*}
and
\begin{eqnarray*}
\underline{U}_2(t,x)=
\begin{cases}
\displaystyle \varepsilon{\psi}(0,x), &t=0,\,\,x\in\overline\Omega, \\[3mm]
\displaystyle \varepsilon{\frac{\kappa_1}{1-\theta}}\psi((\rho\tau)^+,x), &t=(\rho\tau)^+,\,\,x\in\overline\Omega,\\[3mm]
\displaystyle \varepsilon{\frac{\kappa_1}{1-\theta}}e^{-\frac{\mu}{2}t}\psi(t,x),&t\in(0,\rho\tau]\cup((\rho\tau)^+,\tau],\,\,x\in\overline\Omega,
\end{cases}
\end{eqnarray*}
where $(\mu,\phi,\psi)$ is the eigen-pair of problem \eqref{a02} with $\max_{(t,x)\in [0,\tau]\times \overline \Omega}\{\phi(t,x),\psi(t,x)\}= 1$, $\varepsilon>0$ is a sufficiently small constant to be chosen later, and $\kappa_1=(1-\theta)e^{\frac{\mu}{2}\tau }$ is a positive constant to make sure that $\underline{U}_i(0,x)=\underline{U}_i(\tau,x)$ for $i=1,2$. It is  clear that $0<\kappa_1<1-\theta$.

After these, we have
\begin{align*}
\underline {U}_1(0^+,x)-(1+\delta)\underline{U}_1(0,x)
=&\varepsilon{\frac{\kappa_1}{1+\delta}}\phi(0^+,x)-(1+\delta)\varepsilon{\phi}(0,x)\\
=&[\kappa_1-(1+\delta)]\varepsilon\phi(0,x)
<0,
\end{align*}
and
\begin{align*}
\underline {U}_1((\rho\tau)^+,x)-(1-\theta)\underline{U}_1(\rho\tau,x)
=&\varepsilon\kappa_1\phi(\rho\tau,x)-\varepsilon\kappa_1e^{-\frac{\mu}{2}\rho\tau}\phi(\rho\tau,x)\\
=&(1-e^{-\frac{\mu}{2}\rho\tau})\varepsilon\kappa_1\phi(\rho\tau,x)
<0.
\end{align*}
In a similar way, we also have $\underline {U}_2((\rho\tau)^+,x)<(1-\theta)\underline{U}_2(\rho\tau,x)$.

From the definitions, for $t\in(0,\rho\tau]\cup((\rho\tau)^+,\tau]$ and $x\in\overline\Omega$, we obtain
$$\begin{array}{ll}
&\!\!\underline{U}_{1t}-d_1\underline{U}_{1xx}-b(t)\underline{U}_2+(a(t)+m_1(t))\underline{U}_1
+\alpha_1(t)\underline{U}_1^2\\[1mm]
&\leq\varepsilon\frac{\kappa_1}{1-\theta}e^{-\frac{\mu}{2}t}[-\frac{\mu}{2}\phi+\phi_t-d_1\phi_{xx}-
b(t)\psi+(a(t)+m_1(t))\phi+\alpha_1^M\varepsilon\frac{\kappa_1}{1-\theta}e^{-\frac{\mu}{2}t}\phi^2]\\[1mm]
&=\varepsilon\frac{\kappa_1}{1-\theta}\phi(\frac{\mu}{2}+\alpha_1^M\varepsilon\frac{\kappa_1}{1-\theta}e^{-\frac{\mu}{2}t}\phi)\\[1mm]
&\leq0,
\end{array},$$
in a similarly way, we have that
$\underline{U}_{2t}-d_2\underline{U}_{2xx}\leq a(t)\underline{U}_1+m_2(t)\underline{U}_2
+\alpha_2(t)\underline{U}_2^2$ provided that $0<\varepsilon\leq\min\Big\{\frac{-\mu(1-\theta)e^{\frac{\mu}{2}\tau}}{2\alpha_{1}^M\kappa_1},
\frac{-\mu(1-\theta)e^{\frac{\mu}{2}\tau}}{2\alpha_{2}^M\kappa_1}\Big\},$
whence together with the comparison principle follow that $(\underline U_1,\underline U_2)$ is a lower solution to the periodic problem \eqref{c01}.

In the following, the pairs $(\tilde U_1^{(0)},\tilde U_2^{(0)}):=(\overline U_1,\overline U_2)$ and $(\hat U_1^{(0)},\hat U_2^{(0)}):=(\underline U_1,\underline U_2)$ are used to produce two iteration sequences $\{\tilde U_1^{(n)},\tilde U_2^{(n)}\}$ and $\{\hat U_1^{(n)},\hat U_2^{(n)}\}$ that satisfy
\begin{eqnarray}
\left\{
\begin{array}{lll}
\tilde{U}_{it}^{(n)}-d_i\tilde{U}_{ixx}^{(n)}+k_i\tilde{U}_i^{(n)}=F_i(t,\tilde{U}_1^{(n-1)},\tilde{U}_2^{(n-1)}), &t\in(0^+,\rho\tau]\cup((\rho\tau)^+,\tau],\,\,x \in \Omega,  \\[1mm]
\hat{U}_{it}^{(n)}-d_i\hat{U}_{ixx}^{(n)}+k_1\hat{U}_i^{(n)}=F_i(t,\hat{U}_1^{(n-1)},\hat{U}_2^{(n-1)}), &t\in(0^+,\rho\tau]\cup((\rho\tau)^+,\tau]\,\,x \in \Omega,  \\[1mm]
\tilde U_i^{(n)}(t,x)=\hat{U}_i^{(n)}(t,x)=0, & t\in[0,\tau],\,\,x\in\partial\Omega,
\end{array} \right.
\label{c005}
\end{eqnarray}
the periodic conditions for $x\in \overline\Omega$
$$
\tilde U_i^{(n)}(0,x)=\tilde U_i^{(n-1)}(\tau,x),\,\,\hat U_i^{(n)}(0,x)=\hat U_i^{(n-1)}(\tau,x),\ x\in \overline\Omega,\ \,i=1,2,
$$
and the impulsive conditions, for $x\in\Omega$,
$$\tilde U_1^{(n)}(0^+,x)=(1+\delta)\tilde U_1^{(n-1)}(\tau,x),\,\,\hat U_1^{(n)}(0^+,x)=(1+\delta)\hat U_1^{(n-1)}(\tau,x),$$
$$\tilde U_i^{(n)}((\rho\tau)^+,x)=(1-\theta)\tilde U_i^{(n-1)}((\rho+1)\tau,x),\,\,\hat U_i^{(n)}((\rho\tau)^+,x)=(1-\theta)\hat U_i^{(n-1)}((\rho+1)\tau,x)$$
for $i=1,2$, where functions
$F_1(U,V)\triangleq b(t)V-(a(t)+m_1(t))U-\alpha_1(t)U^2+k_1U$ and $F_2(U,V)\triangleq a(t)U-m_2(t)V-\alpha_2(t)V^2+k_2V$ are monotonic increasing in $U$ and $V$ for suitable $k_1$ and $k_2$, respectively.

One checks by \cite{P2005} that
$(\underline U_1,\underline U_2)\leq(\hat U_1^{(m-1)},\hat U_2^{(m-1)})\leq(\hat U_1^{(m)},\hat U_2^{(m)})\leq(\tilde U_1^{(m)},\tilde U_2^{(m)})\leq(\tilde U_1^{(m-1)},\tilde U_2^{(m-1)})\leq(\overline U_1,\overline U_2)$
so that the below limits
$$
\lim\limits_{m\to\infty}{(\hat U_1^{(m)},\hat U_2^{(m)})}=(\hat{U}_1^*,\hat U_2^*) \,{\rm and}\,
\lim\limits_{m\to\infty}{(\tilde U_1^{(m)},\tilde U_2^{(m)})}=(\tilde{U}_1^*,\tilde U_2^*)
$$
are well defined. By letting $n\to\infty$, one yields that $(\hat{U}_1^*,\hat U_2^*)$ and $(\tilde{U}_1^*,\tilde U_2^*)$ are two periodic solutions to the periodic problem \eqref{c01} satisfying
$(\underline U_1,\underline U_2)\leq(\hat{U}_1^*,\hat U_2^*)\leq(\tilde{U}_1^*,\tilde U_2^*)\leq(\overline U_1,\overline U_2).$
Uniqueness of solutions for problem \eqref{c01}, say $(U_1^*(t,x),U_2^*(t,x))$, can be verified as in \cite{XLZ}.

\noindent Item $(ii)$. Initial iterations in $(i)$ assert
$(\hat{U}_1^{(0)},\hat U_2^{(0)})(t,x)\leq{(u_1,u_2)(t,x)}\leq{(\tilde{U}_1^{(0)},\tilde U_2^{(0)})(t,x)}$ for $t\geq{0}$ and $x\in\overline \Omega$,
and the below inequalities
$\hat{U}_i^{(1)}(0,x)=\hat{U}_i^{(0)}(\tau,x)\leq{u_i(\tau,x)}
\leq\tilde{U}_i^{(0)}(\tau,x)=\tilde{U}_i^{(\tau)}(0,x)$ for $x\in\overline \Omega$ and $i=1,2$.

Besides, with the help of problem \eqref{c005}, we obtain for $x\in \overline \Omega$ that
$$\hat{U}_1^{(1)}(0^+,x)=(1+\delta)\hat{U}_1^{(0)}(\tau,x)\leq{(1+\delta) u_1(\tau,x)}
=u_1(\tau^+,x)\leq{(1+\delta) \tilde{U}_1^{(0)}(\tau,x)}=\tilde{U}_1^{(1)}(0^+,x),$$
and
\begin{align*}
\hat{U}_i^{(1)}((\rho\tau)^+,x)=&(1-\theta)\hat{U}_i^{(0)}((\rho+1)\tau,x)\leq{(1-\theta) u_i((\rho+1)\tau,x)}
\\
=& u((\rho\tau)^++\tau,x)\leq{(1-\theta) \tilde{U}_i^{(0)}((\rho+1)\tau,x)} ={\tilde{U}_i^{(1)}((\rho\tau)^+,x)},\,i=1,2.
\end{align*}
Then, comparison argument asserts
 $(\hat{U}_1^{(1)},\hat{U}_2^{(1)})(t,x)\leq{(u_1,u_2)(t+\tau,x)}\leq{(\tilde{U}^{(1)},\tilde{U}^{(1)})(t,x)}$
 for $t\in(0^+,\rho\tau]\cup((\rho\tau)^+,\tau]$ and $x\in \overline \Omega$ so that $(\hat{U}_1^{(1)},\hat{U}_2^{(1)})(t,x)\leq{(u_1,u_2)(t+\tau,x)}\leq{(\tilde{U}_1^{(1)},\tilde{U}_2^{(1)})(t,x)},\,t\geq 0,\,x\in\overline \Omega.$

By iteration, we finally obtain that
$$
(\hat{U}_1^{(m)},\hat{U}_2^{(m)})(t,x)\leq{(u_1,u_2)(t+m\tau,x)}\leq{(\tilde{U}_1^{(m)},\tilde{U}_2^{(m)})(t,x)},\,t\geq{0},\,x\in\overline \Omega
$$
holds for any $m$, whence follows together with the uniqueness of the solution to problem \eqref{c01} that $\lim\limits_{m\to\infty}{(\hat{U}_1^{(m)},\hat{U}_2^{(m)})(t,x)}=
\lim\limits_{m\to\infty}{(\tilde{U}_1^{(m)},\tilde{U}_2^{(m)})(t,x)}=(U_1^*,U_2^*)(t,x).$ This ends the proof.
\epf

\section{Numerical simulations}
In this section, the numerical approximation will be used to verify the analytical findings. We fix some parameters $d_1=d_2=0.01$, $b(t)=a(t)=0.8$, $m_1(t)=0.1$, $m_2(t)=0.5$, $\alpha_1(t)=0.1$, $\alpha_2(t)=0.2$ and $\rho=0.5.$

\begin{figure}[!ht]
\centering
\subfigure[]{ {
\includegraphics[width=0.28\textwidth]{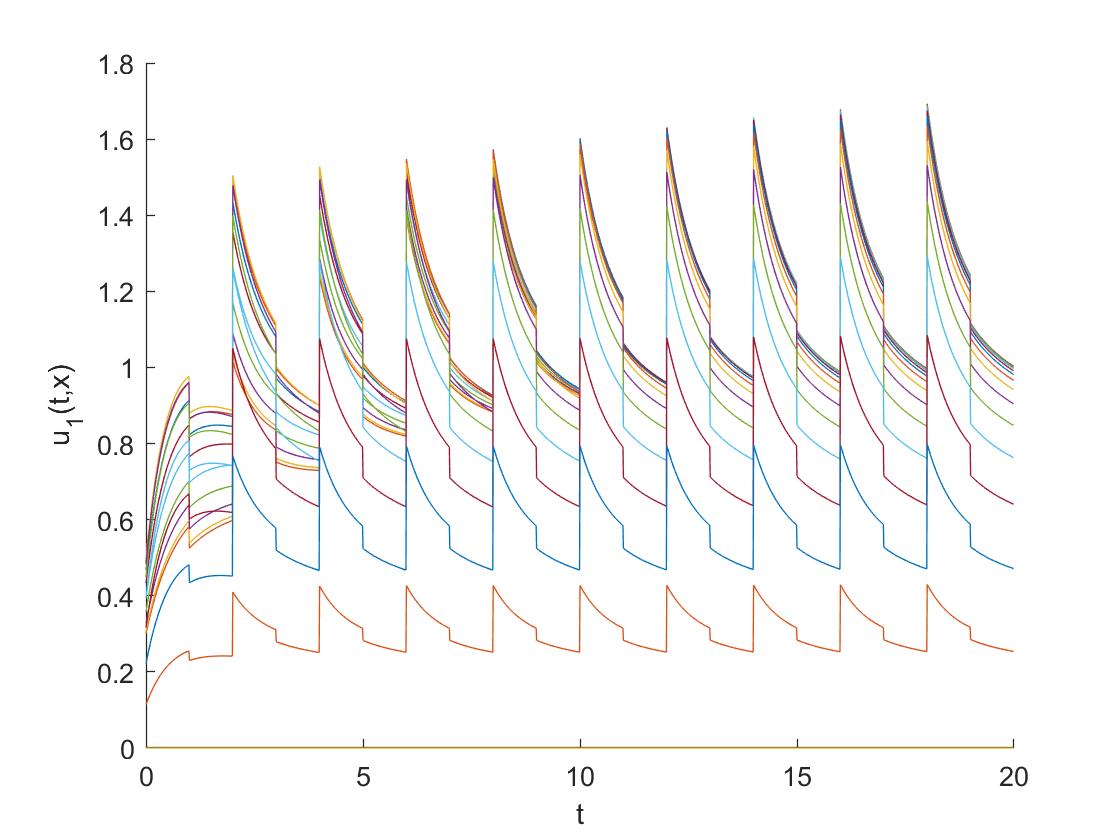}
} }
\subfigure[]{ {
\includegraphics[width=0.28\textwidth]{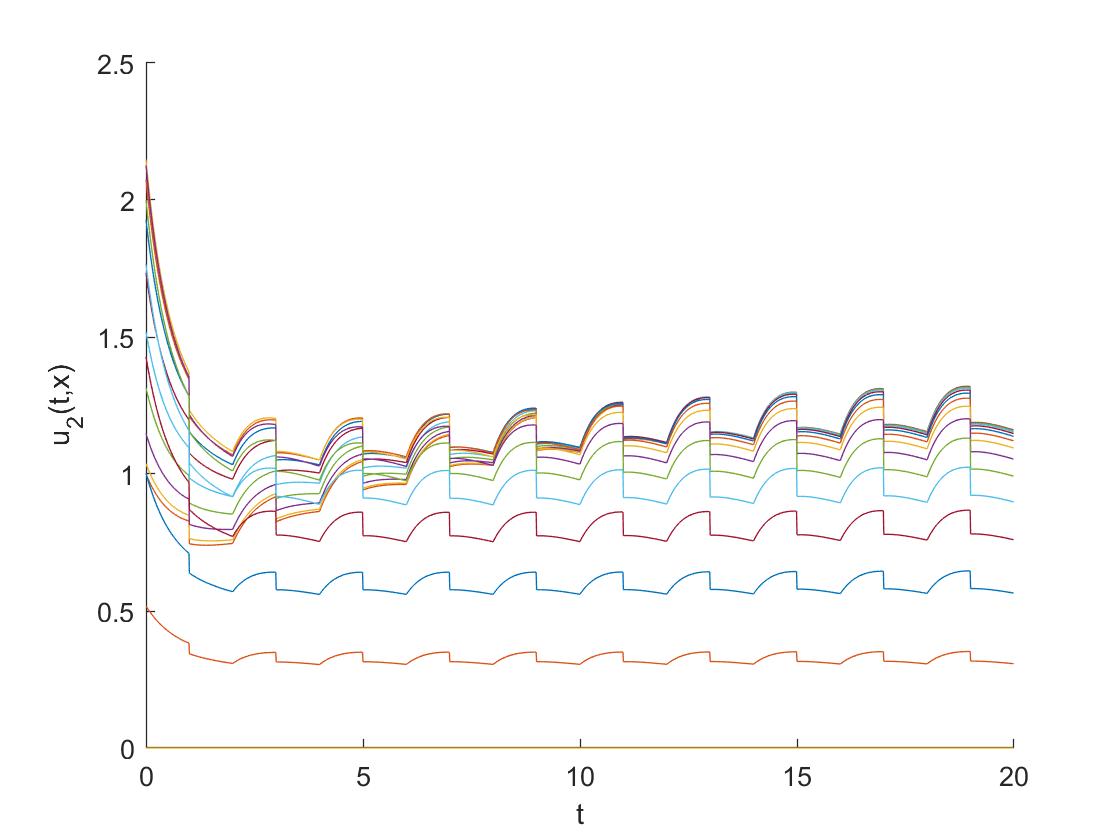}
} }
\subfigure[]{ {
\includegraphics[width=0.28\textwidth]{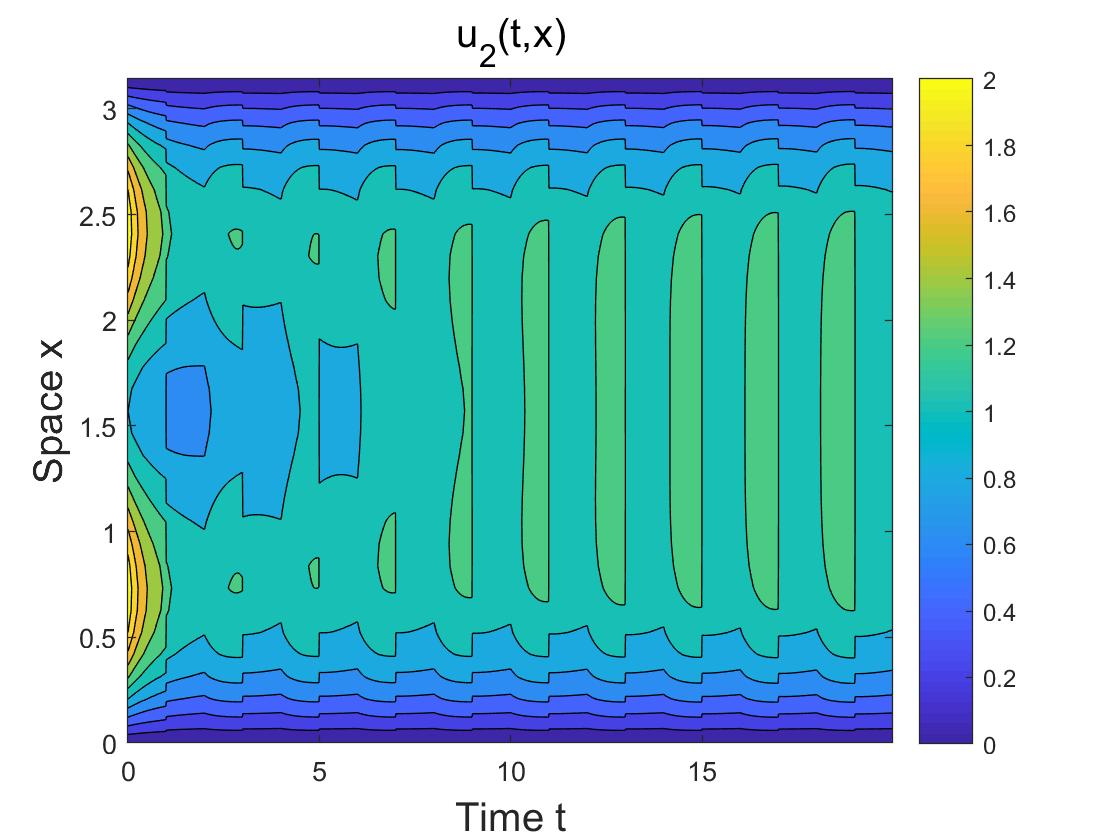}
} }
\caption{\scriptsize Simulations with: $\delta=0.7$, $\theta=0.1$ and $\tau=2$. Birth pulses happen at every time $t=0,2,4,\dots$ in $u_1$ and harvesting pulses occur at every time $t=1,3,5,\dots$ in $u_1$ and $u_2$. Graphs (a) and (b) are the projection of $u_1$ and $u_2$ in $t$ plant, respectively, Graph (c) is the cross-section view of $u_2$ in $t-x$ plant. It shows that juveniles and adults will spatially spread.
}
\label{tu1}
\end{figure}

\begin{figure}[!ht]
\centering
\subfigure[]{ {
\includegraphics[width=0.28\textwidth]{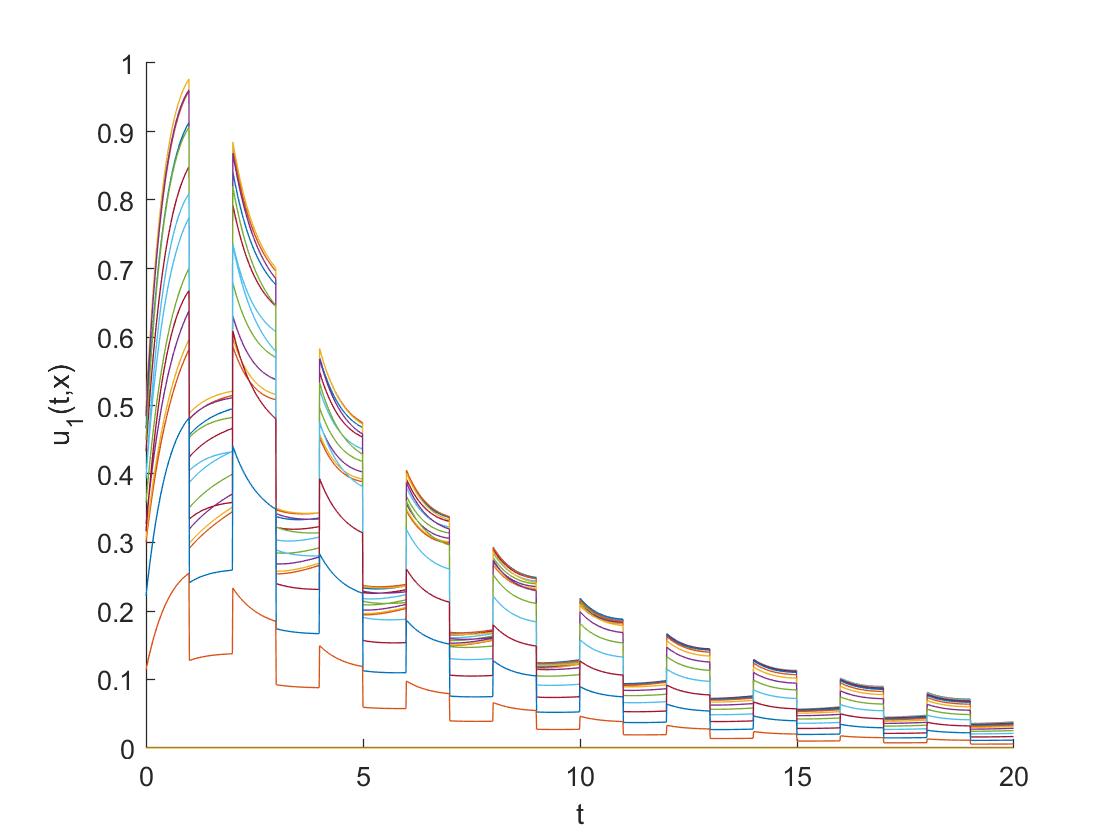}
} }
\subfigure[]{ {
\includegraphics[width=0.28\textwidth]{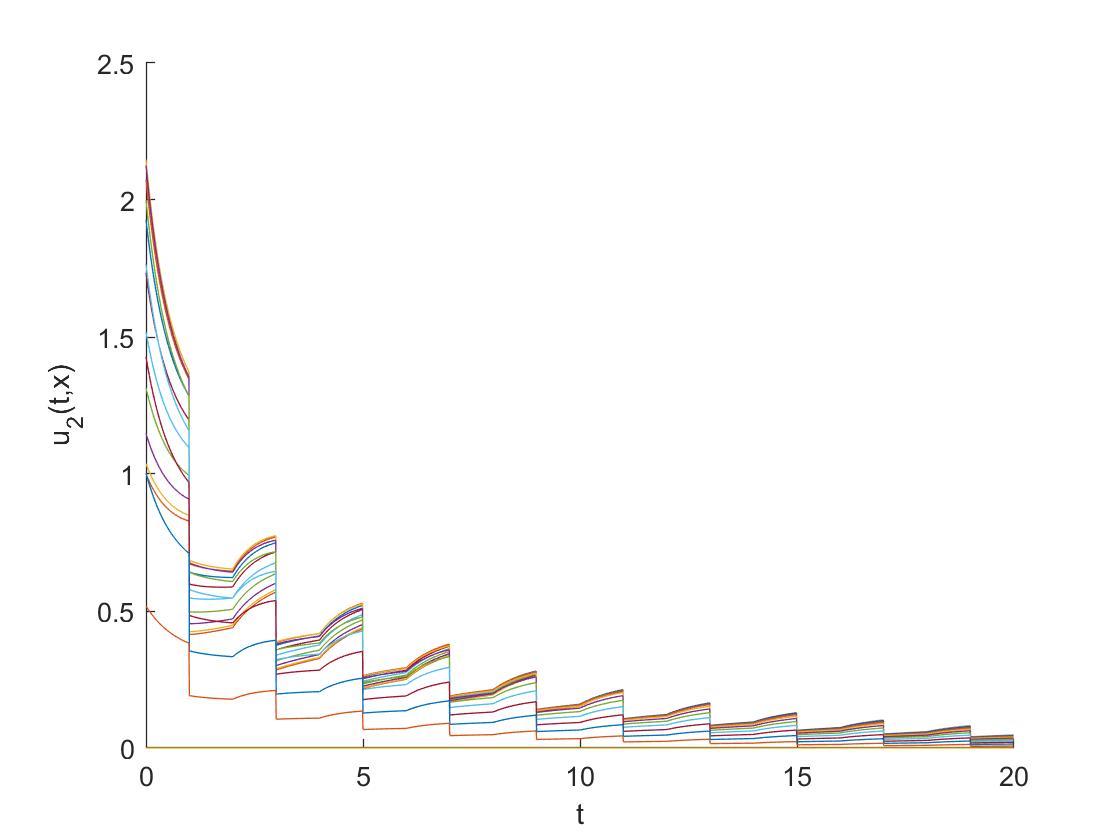}
} }
\subfigure[]{ {
\includegraphics[width=0.28\textwidth]{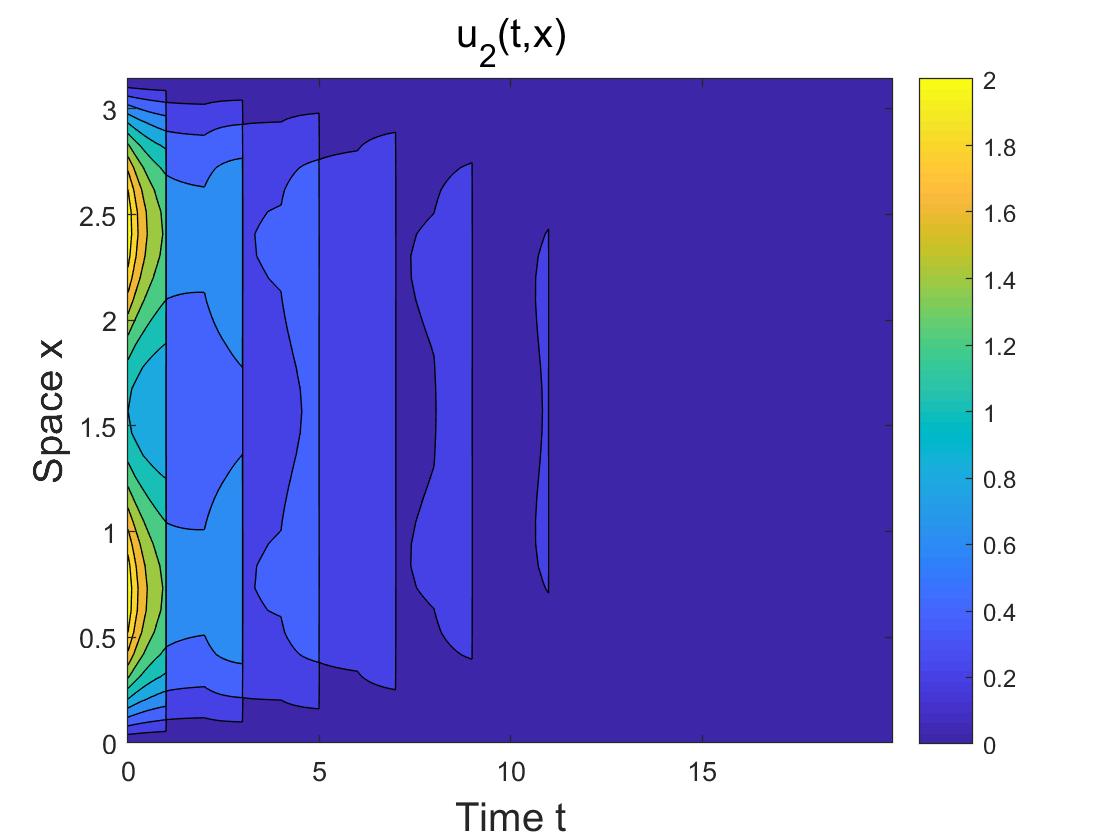}
} }
\caption{\scriptsize  Simulations with: $\delta=0.7$, $\theta=0.5$ and $\tau=2$. Graph (a) shows that birth pulse happen at time $t=0,2,4,\dots$ and harvesting pulse occur at time $t=1,3,5,\dots$. Graph (b) describes only impulsive harvesting occurs. Species eventually vanish.
}
\label{tu2}
\end{figure}

To consider the impact of harvesting intensity $\theta$, we firstly fix $\tau=2$ and $\delta=0.7$ in Figs \ref{tu1} and \ref{tu2}. In this case, birth pulses happen at $t=0,2,4,\dots$ in juveniles and harvesting pulses occur at $t=1,3,5,\dots$ in juveniles and adults. We now choose $\theta=0.1$ in Fig. \ref{tu1} and $\theta=0.5$ in Fig. \ref{tu2} with other parameters unchanged, one easily sees that a larger harvesting intensity $\theta$ makes species from spread in Fig. \ref{tu1} to vanish in Fig. \ref{tu2}.

\begin{figure}[!ht]
\centering
\subfigure[]{ {
\includegraphics[width=0.28\textwidth]{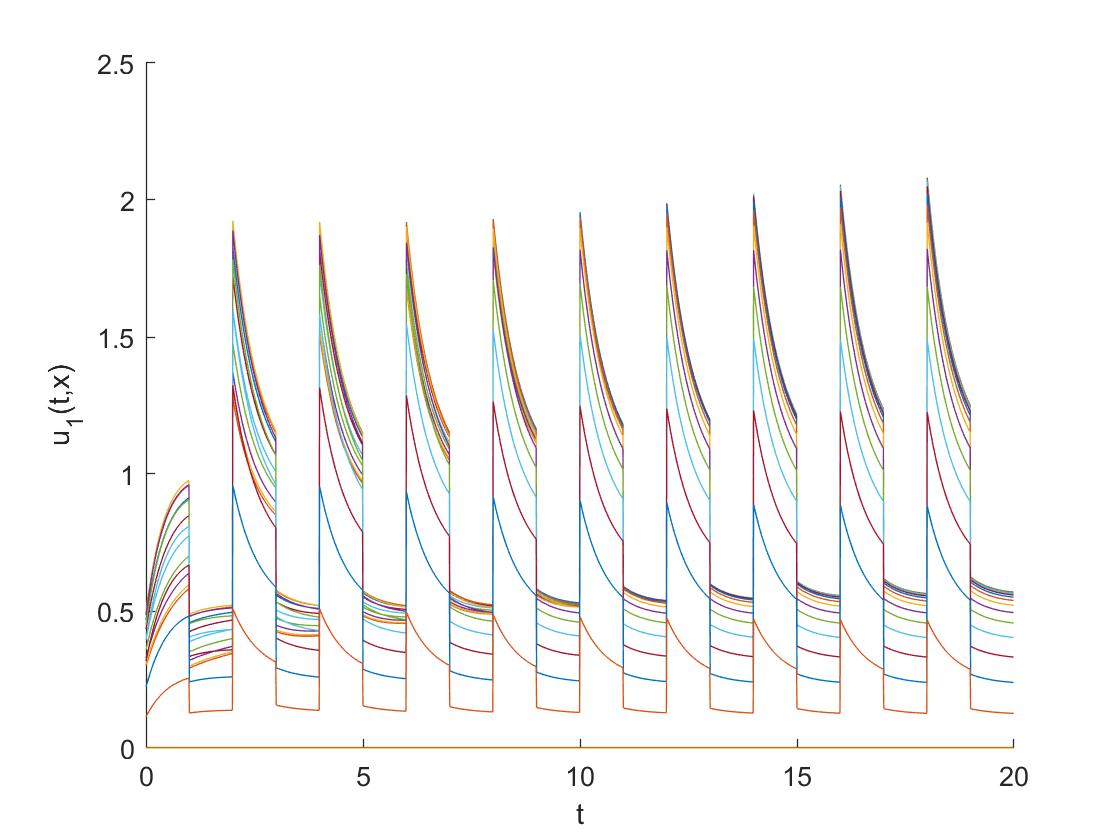}
} }
\subfigure[]{ {
\includegraphics[width=0.28\textwidth]{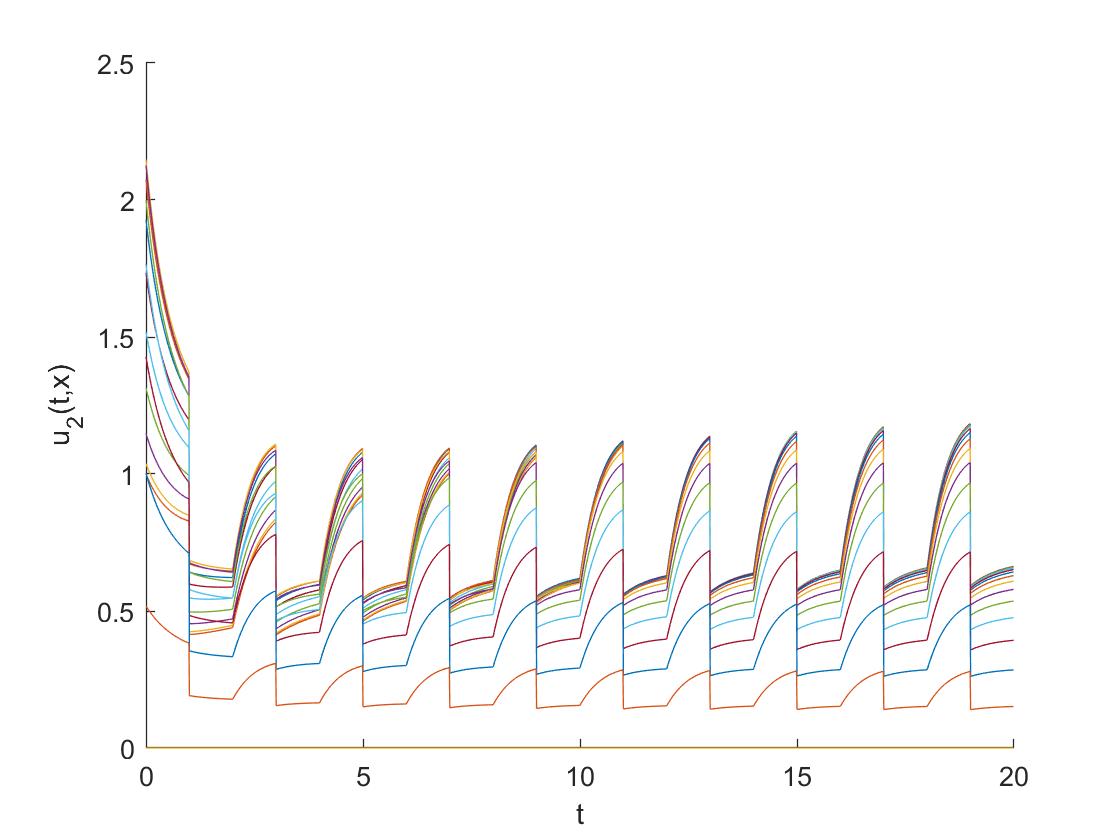}
} }
\subfigure[]{ {
\includegraphics[width=0.28\textwidth]{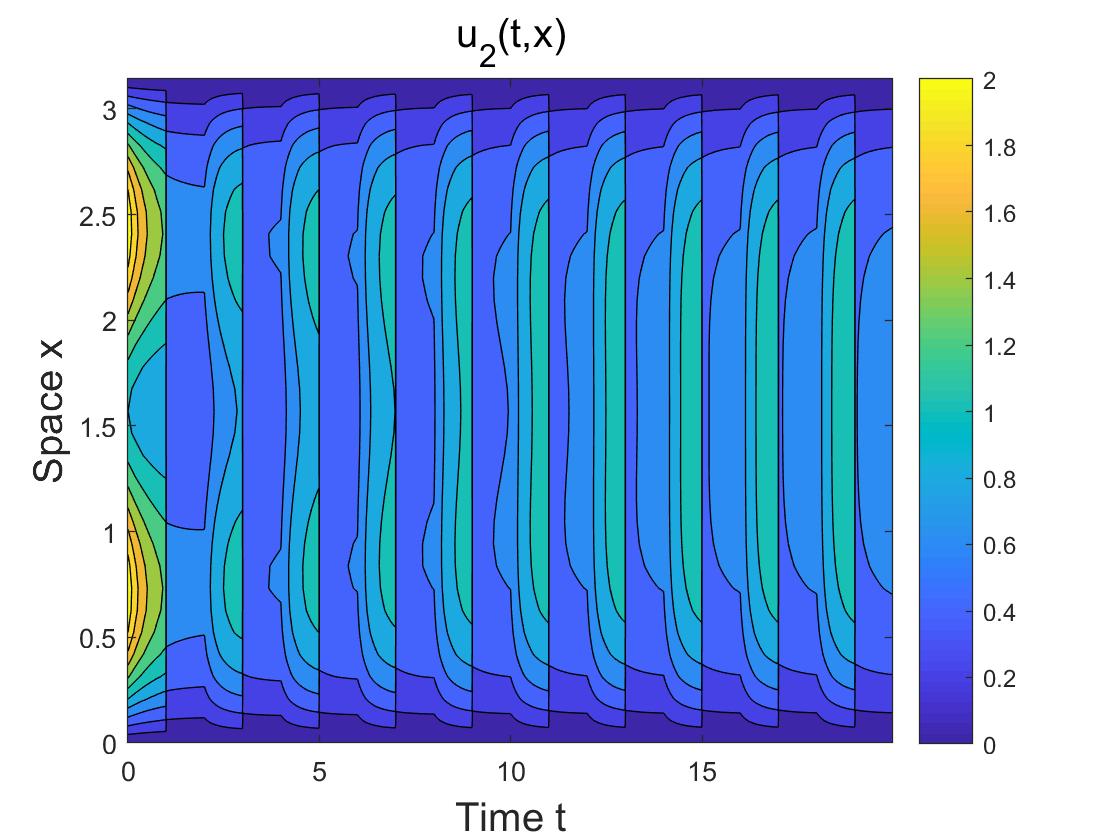}
} }
\caption{\scriptsize  Simulations with: $\delta=2.7$, $\theta=0.5$ and $\tau=2$. Graphs (a)-(c) imply that juveniles and adults stabilized  to a positive steady state.
}
\label{tu3}
\end{figure}

Next, we fix $\tau=2$ and $\theta=0.5$ in Figs \ref{tu2} and \ref{tu3}. The impact of birth intensity $\delta$ is considered by choosing different $\delta$ and keeping other parameters unchanged. In  comparison to Fig. \ref{tu2} with $\delta=0.7$ and Fig. \ref{tu3} with $\delta=2.7$,  we observe that when birth perturbation occurs, large birth intensity $\delta$ makes species from vanish in Fig. \ref{tu2} to persist in Fig. \ref{tu3}. So large birth intensity is conductive to the existence of fish.

\newpage

\begin{figure}[!ht]
\centering
\subfigure[]{ {
\includegraphics[width=0.28\textwidth]{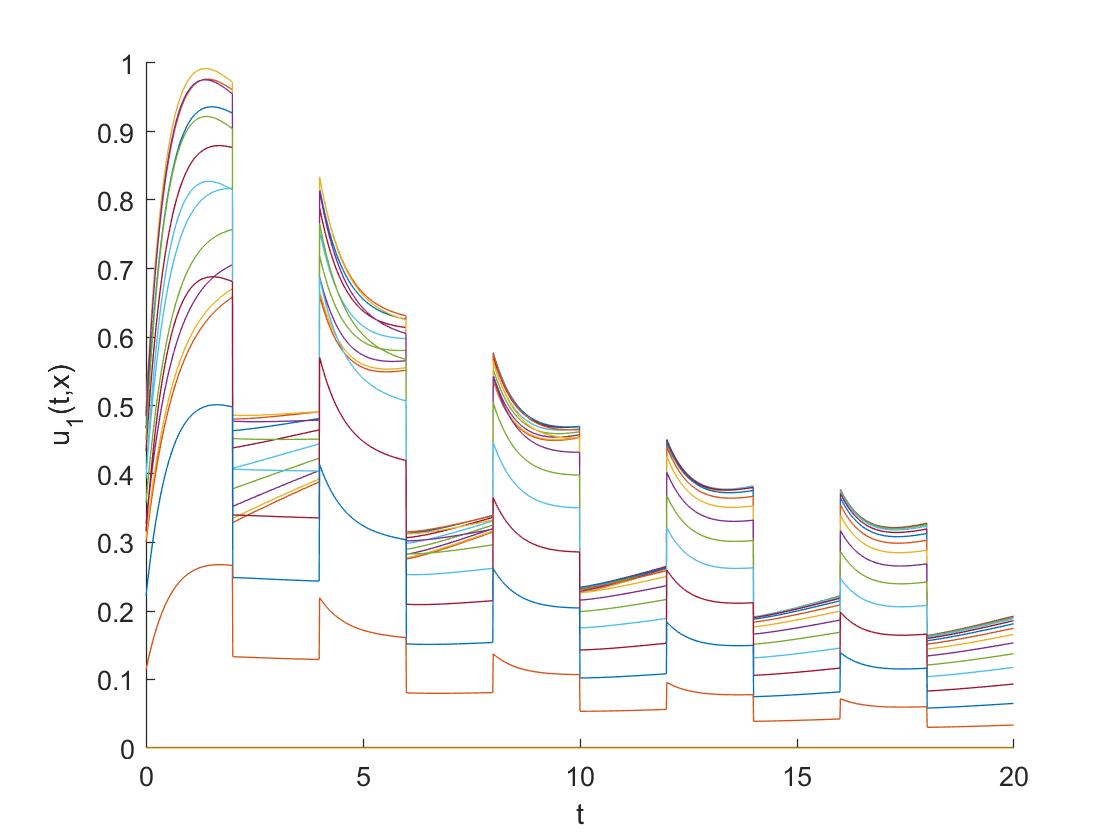}
} }
\subfigure[]{ {
\includegraphics[width=0.28\textwidth]{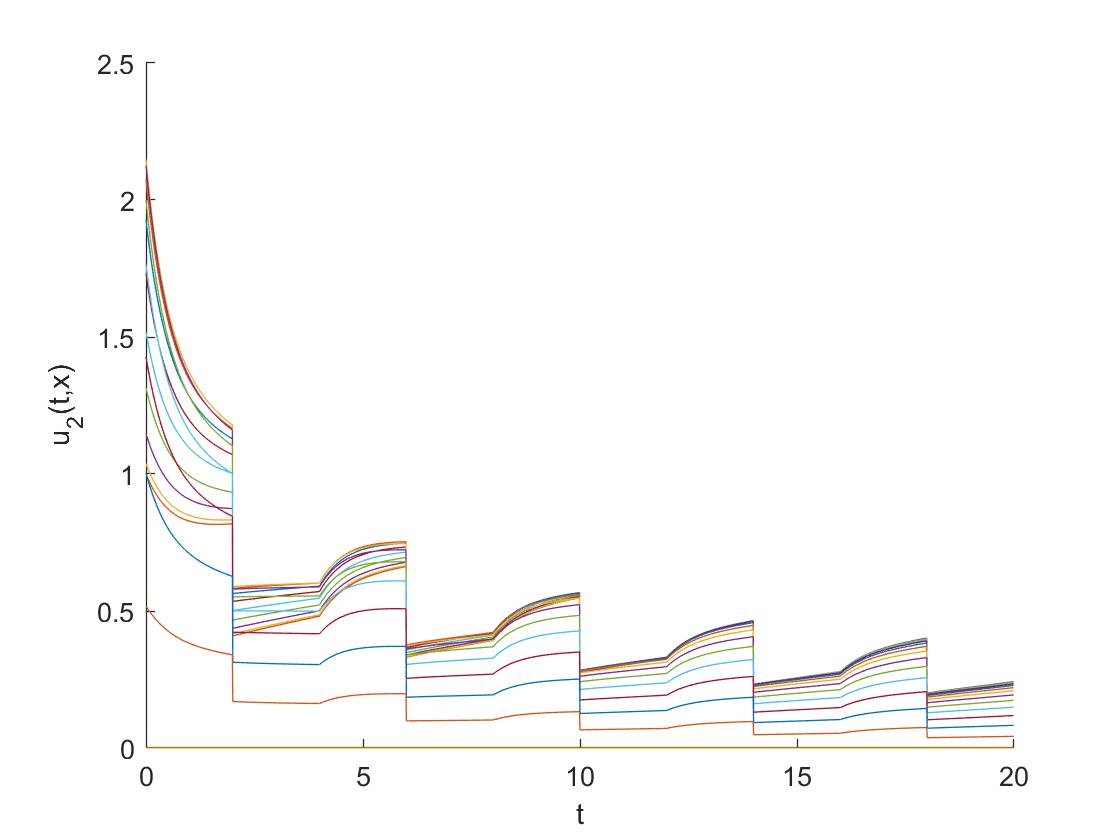}
} }
\subfigure[]{ {
\includegraphics[width=0.28\textwidth]{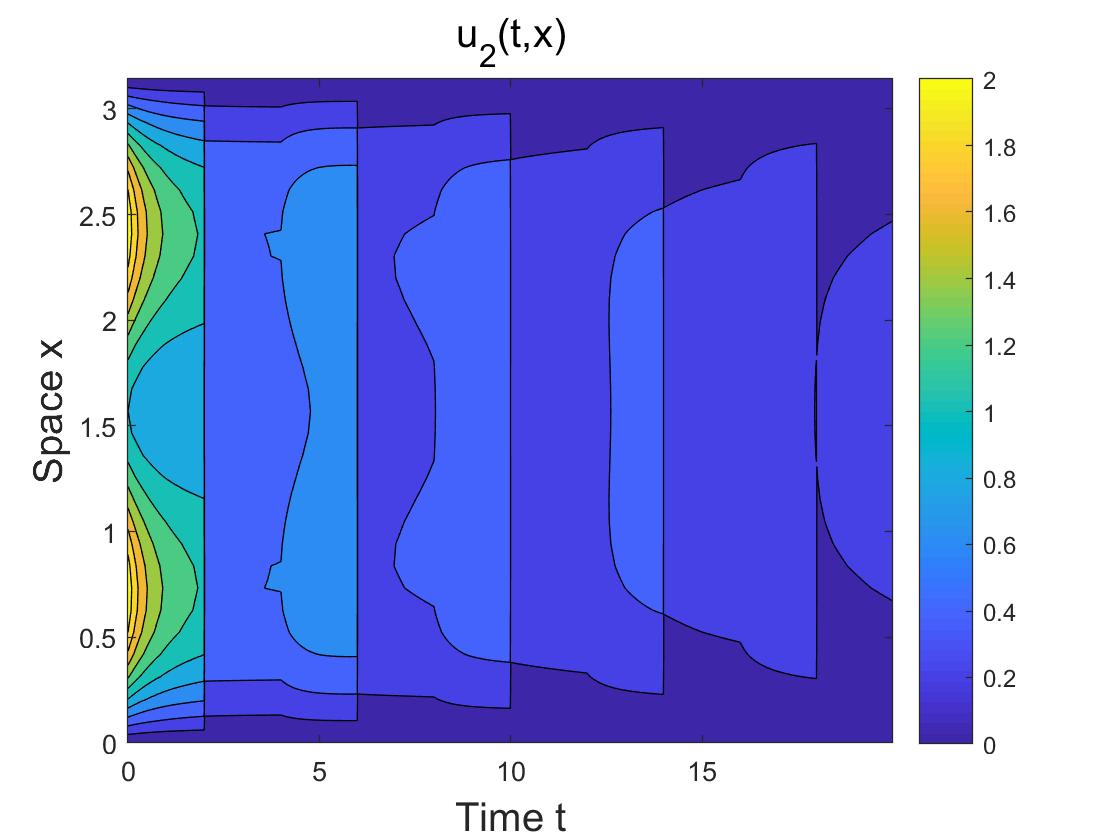}
} }
\caption{\scriptsize  Simulations for $\tau=4$ and other parameters are the same as in Fig. \ref{tu2}. At every time $t=0,4,8,\dots$, birth pulses happen in juveniles and at every time $t=2,6,10,\dots$, harvesting pulses appear in juveniles and adults. Species finally converge to a positive steady state.
}
\label{tu41}
\end{figure}

The numerical simulations in the above illustrate pulse intensity on species spreading and vanishing. Whether does the timing of pulse affect the long-time behavior of species? In the following,  $\tau=2$ in Fig. \ref{tu2} is replaced by $\tau=4$ in Fig. \ref{tu41}. Then birth pulses happens at every time $t=0,2,4,\dots$ in Fig. \ref{tu2}, but happens at $t=0,4,8,\dots$ in Fig. \ref{tu41}, and harvesting pulses happens at $t=1,3,5,\dots$ in Fig. \ref{tu2}, and in Fig. \ref{tu41}, it happens at $t=2,6,10,\dots$. It follows from Figs. \ref{tu2} ($\tau=2$ for extinction) and \ref{tu41} ($\tau=4$ for persistence) that the change of pulses timing contributes to the change of steady state.

\vspace{1mm}
To sum up, we propose an age-structured model with  birth pulse exacting on juveniles and harvesting pulse acting on the whole individuals. The global existence, uniqueness and estimates of solution to problem \eqref{a01} were given in Section 2. Three ways  to determine the principal eigenvalue with multiple pulses in different cases were presented in Section 3.

 Properties of the principal eigenvalue in the birth intensity $\delta$, harvesting intensity $\theta$ and the habitat $\Omega$
were analysed in Theorem \ref{5c01}, and some dynamic behaviors related to pulses were investigated, see vanishing of fish in Theorem \ref{ab} for the case $\mu(\delta,\theta,\Omega)>0$ and spreading in Theorem \ref{ad} for the case $\mu(\delta,\theta,\Omega)<0$.

Theoretical results have been verified by numerical approximations, and specially,  the impact of intensity of harvesting ($\theta$), birth ($\delta$) and timing of pulse ($\tau$) have been emphasized. Recalling Theorems \ref{5c01} $(ii)$ and \ref{ab}, we can see that the larger the harvesting intensity ($\theta$), the more unfavorable for spreading of species, which coincides with Figs. \ref{tu1} and \ref{tu2}. Also, Figs. \ref{tu2} and \ref{tu3} show that large birth intensities ($\delta$) are benefit for the species spreading. It should be mentioned that  the timing of pulse ($\tau$) can also result in the changing of  the existing status,  see vanishing in Fig. \ref{tu2} ($\tau=2$) and spreading in Fig. \ref{tu41} ($\tau=4$) with other parameters fixed. Whether does the optimal timing of pulses exist and how to choose it for sustainable development of fishery resources? All of these require further study.

 {\bf Declaration of Competing Interest:}\ Author(s) have no conflict of interest.

 {\bf Data Availability:}\ The author confirms that the data supporting the findings of this study are
available within the manuscript.

\end{document}